{

\documentclass[12pt,a4paper]{article}
\usepackage{graphicx}
\usepackage[T1]{fontenc}
\usepackage[utf8]{inputenc}
\usepackage{textcomp}
\usepackage[sc,osf]{mathpazo}
\usepackage{a4wide}  
\usepackage{latexsym,amsthm,amsfonts,amsmath,mathrsfs,amssymb}
\usepackage{dsfont}
\usepackage{accents}
\usepackage[nosort]{cite}
\usepackage{booktabs} 
\usepackage[unicode,implicit]{hyperref}
\hypersetup{%
  pdftitle    = {Extremal stringy black holes in equilibrium at first order in $\alpha'$}
  pdfkeywords = {black holes, stringy black holes, extremal black holes,
    multi-center, equilibrium,supersymmetry,
    alpha prime corrections,string theory,heterotic superstring, higher-order corrections,stability},
  pdfauthor   = {Tom\'as Ort\'{\i}n, Alejandro Ruip\'erez and Matteo Zatti},
  plainpages  = true,
  colorlinks  = true,
  citecolor   = blue,
  urlcolor    = red,
  linkcolor   = black
}
\newcommand{\hepth}[1]{{\tt
\href{http://www.arXiv.org/abs/hep-th/#1}{hep-th/#1}}}
\newcommand{\grqc}[1]{{\tt
\href{http://www.arXiv.org/abs/gr-qc/#1}{gr-qc/#1}}}

\newcommand{\arxiv}[1]{{\tt arXiv:\href{http://www.arXiv.org/abs/#1}{#1}}}
\newcommand{\doi}[1]{{\tt DOI:\href{http://dx.doi.org/#1}{#1}}}

\makeatletter
\@addtoreset{equation}{section}
\makeatother

\pagestyle{empty}

\begin{document}

\begin{flushright}
\small
IFT-UAM/CSIC-21-066\\
December 23\textsuperscript{rd}, 2021\\
Revised March 30\textsuperscript{th}, 2023\\
\normalsize
\end{flushright}

\vspace{1cm}

\begin{center}

  {\Large {\bf Extremal stringy black holes in equilibrium\\[.5cm]
      at first order in $\alpha'$}}

\vspace{1.5cm}

\renewcommand{\thefootnote}{\alph{footnote}}

{\sl\large  Tom\'{a}s Ort\'{\i}n,}$^{1,}$\footnote{Email: {\tt tomas.ortin[at]csic.es}}
{\sl\large Alejandro Ruip\'erez$^{2,3,}$}\footnote{Email: {\tt alejandro.ruiperez[at]pd.infn.it}}
{\sl\large and Matteo Zatti$^{1,}$}\footnote{Email: {\tt matteo.zatti[at]estudiante.uam.es}}

\setcounter{footnote}{0}
\renewcommand{\thefootnote}{\arabic{footnote}}
\vspace{1cm}

${}^{1}${\it Instituto de F\'{\i}sica Te\'orica UAM/CSIC\\
C/ Nicol\'as Cabrera, 13--15,  C.U.~Cantoblanco, E-28049 Madrid, Spain}

\vspace{0.5cm}

${}^{2}${\it Dipartimento di Fisica ed Astronomia ``Galileo Galilei'',\\
  Universit\`a di Padova, Via Marzolo 8, 35131 Padova, Italy}

\vspace{0.5cm}

${}^{3}${\it INFN, Sezione di Padova,Via Marzolo 8, 35131 Padova, Italy}

\vspace{1.5cm}


{\bf Abstract}
\end{center}
\begin{quotation}
  {\small We compute the first-order $\alpha'$ corrections to well-known
    families of heterotic multi-center black-hole solutions in five and four
    dimensions. The solutions can be either supersymmetric or
    non-supersymmetric, depending on the relative sign between two of the
    black-hole charges. For both cases, we find that the equilibrium of forces
    persists after including the $\alpha'$ corrections, as the existence of
    multi-center solutions free of unphysical singularities shows. We analyze
    the possibility of black-hole fragmentation.
  }
\end{quotation}

\newpage
\pagestyle{plain}

\tableofcontents

\newpage

\section{Introduction}

Over the last 30 years many extremal and multi-center black-hole solutions of
supergravity theories have been obtained, supersymmetric and
non-supersymmetric.\footnote{For a review on black-hole solutions with many
  references see, for instance, \cite{Ortin:2015hya}.} Many of those theories
can also be viewed as low-energy effective field theories of different string
theory compactifications to lowest order in $\alpha'$ and in the string
coupling constant. This allowed Strominger and Vafa to establish a relation
between the entropy of the corresponding solutions (\textit{stringy black-hole
  solutions}) and the microscopic entropy of the corresponding
compactification of string theory which, in general, includes branes and other
extended objects \cite{Strominger:1996sh}.

It is natural to try to extend this relation between macroscopic and
microscopic entropies to more general solutions and, more importantly perhaps,
to higher orders in $\alpha'$ and in the string coupling constant. The
calculations of both the macroscopic and microscopic entropies have to be
improved to the next order, independently, to be later compared. If this
comparison is to be understood as an unbiased test of this relation and of
the viability of String Theory as a theory of Quantum Gravity, each of these
separate calculations must be internally self-consistent: the correctness of
one of them cannot solely rely on its agreement with the other one if we do
not want to suffer confirmation bias.

In the macroscopic side, this suggests the following program:

\begin{enumerate}
\item Find the first corrections in $\alpha'$ and in the string coupling
  constant to the relevant superstring effective field theory.
\item Find single- and multi-center black-hole solutions to the corrected
  action. This can be done perturbatively so that the known zeroth-order
  solutions are recovered when $\alpha'$ is set to zero.
\item Compute the macroscopic entropy associated to the corrected black hole
  solution. This entropy will no longer be the Bekenstein-Hawking one, since
  the corrected action will contain higher-curvature terms and can no longer
  be understood as General Relativity coupled to a prescribed set of
  interacting matter. One has to use Wald's entropy.
\item The macroscopic entropy must be written in terms of charges that can be
  unambiguously related to a string background, \textit{i.e.}~charges that can
  be associated to numbers of branes. 
\end{enumerate}

Only after a satisfactory completion of this program can the macroscopic
entropy be compared to the microscopic one, which must have been computed
independently.

This program, though, presents many practical and conceptual difficulties:

\begin{enumerate}
\item We have a very limited knowledge of the corrections to the zeroth order
  superstring effective action, specially of the corrections in the string
  coupling constant, which are mainly unknown.

  We know that the first corrections in $\alpha'$ start at third order for the
  type~II superstring theories, but we do not have a complete knowledge of the
  effective action to that order. This prevents us from doing fully reliable
  calculations of $\alpha'$ corrections to the solutions of these theories.

  The situation with the heterotic superstring is much better: we know in full
  detail the $\alpha'$ corrections to the effective action up to third order
  and in a form consistent with local spacetime supersymmetry
  \cite{Bergshoeff:1989de}.\footnote{See also earlier work in
    Refs.~\cite{Gross:1986mw,Metsaev:1987zx}.} In contrast, there is very
  limited information about the string loop corrections
  \cite{Chemissany:2007he}. 

\item Given the limitations in our knowledge of the string effective action we
  have just discussed, it is natural to focus on trying to obtain the first
  $\alpha'$ corrections to black-hole solutions of the heterotic superstring
  imposing the necessary conditions on the parameters to ensure that the
  string coupling constant is small enough that the string loop corrections
  can be safely neglected.\footnote{Actually, these and other similar
    conditions that guarantee that the $\alpha'$ corrections are also
    negligible must also be imposed on the zeroth-order solutions as well.}

  Furthermore, given the complexity of the corrected equations of motion,
  it is also natural to consider at first, single, static,
  spherically-symmetric, extremal black holes with the minimal number of
  charges that produce a regular event horizon and start by computing the
  first-order $\alpha'$ corrections only. These are the so-called 3-charge and
  4-charge black holes in 5 and 4 dimensions, respectively and all of them can
  be embedded in the heterotic superstring effective action. The 3-charge ones
  include a black hole dual to the one originally considered by
  Strominger-Vafa \cite{Strominger:1996sh}, which is supersymmetric, together
  with similar, non-supersymmetric black holes. The 4-charge ones include the
  heterotic version of the Maldacena-Strominger-Johnson-Khuri-Myers black hole
  \cite{Maldacena:1996gb,Johnson:1996ga}, which is supersymmetric, plus
  several similar non-supersymmetric black holes.\footnote{An exhaustive study
    of the different non-supersymmetric 4-dimensional 4-charge back holes can
    be found in Ref.~\cite{Cano:2021nzo}.}

  The first-order $\alpha'$ corrections to all these simplest but paradigmatic
  extremal black-hole solutions have been found quite recently in a series of
  papers
  \cite{Cano:2018qev,Chimento:2018kop,Cano:2018brq,Cano:2019oma,Cano:2021nzo}
  and now it is natural to take the next step and consider more complicated
  solutions: non-extremal, multi-center, stationary...

  There are good reasons to consider first, after the extremal 3- and 4-charge
  solutions, the non-extremal ones, as we are going to argue when we discuss
  the calculation of the macroscopic entropy.  The first-order $\alpha'$
  corrections to some non-extremal solutions have also been found
  \cite{Cano:2019ycn,Cano:2022tmn}, but they are much harder to obtain since
  the general results proven in \cite{Chimento:2018kop} cannot be applied to
  the non-extremal solutions. Thus, it will be long before the corrections to
  the non-extremal 4-charge black holes will be found.\footnote{In
    Ref.~\cite{Zatti:2023oiq} the first-order $\alpha'$ corrections to the
    4-charge non-extremal solutions have been computed, but only in the
    particular case in which two of them are equal. The case in which the 4
    charges are independent still seems out of our reach.}

  The general results proven in \cite{Chimento:2018kop} do not seem to apply
  to stationary solutions either but, fortunately for us, they do apply with
  little modifications for the multi-center ones, and, in this paper we are
  going to profit from them to construct the first $\alpha'$-corrected
  multi-center black-hole solutions both in 5- and in 4-dimensions and both
  supersymmetric and non-supersymmetric. They are the multi-center
  generalization of the single, static, extremal black-hole solutions whose
  $\alpha'$-corrections were found in
  Refs.~\cite{Cano:2018qev,Chimento:2018kop,Cano:2018brq,Cano:2019oma,Cano:2021nzo}
  and include most of them.\footnote{We have not found the multi-center
    generalization of all the possible non-supersymmetric black holes
    considered in Ref.~\cite{Cano:2021nzo}.}  Before we introduce this
  particular work, though, it is convenient to discuss the rest of the program
  outlined above since it affects the reliability of our results.

  It is worth mentioning, though, that the solutions presented in
  Refs.~\cite{Cano:2018qev,Chimento:2018kop,Cano:2018brq,Cano:2019oma,Cano:2021nzo}
  satisfy a highly non-trivial consistency test: they are invariant (as
  families of solutions) under the $\alpha'$-corrected Buscher T~duality
  transformations derived in Ref.~\cite{Bergshoeff:1995cg} (see also
  Ref.~\cite{Elgood:2020xwu}). The multi-center solutions that we are going to
  present, being related to the general solutions of
  Ref.~\cite{Chimento:2018kop} pass this strict test as well.

\item At lowest order in $\alpha'$ and the string coupling constant, all
  string effective field theories can be regarded as General Relativity (GR)
  coupled to particular kinds of interacting matter and the entropy of any of
  their black-hole solutions is the Bekenstein-Hawking entropy: one quarter of
  the area of the horizon in Planck units. This relation originates in the
  proofs of the first and second laws of black hole mechanics
  \cite{Hawking:1971vc,Bardeen:1973gs} (based on Einstein's equations)
  together with the discovery of the Hawking radiation \cite{Hawking:1974sw}
  and the precise relation between its (Hawking) temperature and the surface
  gravity of the black hole.

  At higher orders in $\alpha'$, string effective actions contain terms of
  higher order in the curvatures and can no longer be regarded as GR. However,
  we know that their black-hole solutions still behave as thermal objects
  since the phenomenon of Hawking radiation does not depend on the dynamics of
  gravity. Furthermore, it can be shown that the zeroth law (the fact that
  surface gravity is constant over the horizon) \cite{Bardeen:1973gs} can be
  proven without the use of the Einstein equations \cite{Racz:1995nh} and,
  therefore, it will still hold in the $\alpha'$-corrected string effective
  field theory actions.

  In GR, the relation between the area and the entropy was hinted at by the
  results obtained by Christodoulou and
  Hawking~\cite{Christodoulou:1970wf,Christodoulou:1972kt,Hawking:1971tu} but
  there are no analogous results for other theories of gravity suggesting a
  candidate to be identified with the entropy. For this reason, Wald's
  derivation of the first law of black hole mechanics for arbitrary
  diff-invariant theories of pure gravity \cite{Wald:1993nt}, based on
  previous work with Lee \cite{Lee:1990nz}, constituted a great breakthrough
  in this field. The quantity whose variation with respect to the mass is the
  inverse of the Hawking temperature must be the black hole entropy and, in
  this framework, it is known as Wald entropy. In order to confirm this
  identification one has to prove that this quantity also satisfies the second
  law. This is a much harder problem, but there is real progress in this
  direction \cite{Wall:2015raa,Hollands:2022fkn,Davies:2022xdq}.

  As we have mentioned, the string effective action is a theory of gravity
  coupled to very special interacting matter. Thus, an extension of Wald's
  results including matter is necessary to deal with it. In
  Ref.~\cite{Iyer:1994ys}, Iyer and Wald considered diff-invariant theories of
  gravity coupled to matter fields transforming as tensors under
  diffeomorphisms, obtaining, as a result, the celebrated Iyer-Wald
  prescription to derive the entropy formula which has been widely used to
  compute the Wald entropy of $\alpha'$-corrected near-horizon limits of
  stringy black holes (see,
  \textit{e.g.}~Refs.~\cite{Sen:2005iz,DominisPrester:2008ynb} and references
  therein).

  This formula was used in
  Refs.~\cite{Cano:2018qev,Chimento:2018kop,Cano:2018brq,Cano:2019oma} to
  compute the Wald entropy of the extremal black hole solutions found in those
  same references, but it was soon realized that the Iyer-Wald prescription
  cannot be directly and reliably applied to the heterotic superstring
  effective action:

  \begin{enumerate}
  \item The entropy formula produced is different if one uses the
    10-dimensional or the toroidally-compactified actions
    \cite{Faedo:2019xii}.
  \item The entropy formula produced is Lorentz frame-dependent
    \cite{Elgood:2020nls}, which is utterly unacceptable. 
  \end{enumerate}
  
  Both problems arise from the inadequate treatment of the
  Lorentz-Chern-Simons terms present in the Kalb-Ramond field strength in the
  heterotic superstring effective action at first order in
  $\alpha'$.\footnote{It should be stressed that these Chern-Simons terms are
    very different from those considered in Ref.~\cite{Tachikawa:2006sz},
    present in the action and introducing additional total derivatives in its
    variation under diffeomorphisms: they transform as 3-forms under
    diffeomorphisms, to start with.} However, ultimately, this inadequacy
  follows from the over-restrictive assumptions made on the matter fields in
  Ref.~\cite{Iyer:1994ys}, which leads to the total absence of
  non-gravitational work terms in the first law of black hole mechanics, such
  as those proportional to the variations of electric charges. As explained in
  \cite{Elgood:2020svt}, for instance, most matter fields are not simple
  tensor fields under diffeomorphisms, but sections of principal bundles or
  more complicated structures. The only true tensor field in the Standard
  Model is the metric.\footnote{See Ref.~\cite{Prabhu:2015vua} for a rigorous
    treatment of the principal bundle case and Ref.~\cite{Hajian:2015xlp} for
    a different take on this problem.}
  
  In Refs.~\cite{Elgood:2020svt,Elgood:2020mdx} we revised from this point of
  view Wald's derivation of the first law of black hole mechanics in simple
  theories, obtaining the expected work terms and, as a bonus, restricted
  forms of the generalized zeroth laws.\footnote{Again, in the principal
    bundle case, this result was obtained in Ref.~\cite{Prabhu:2015vua}.} In
  Ref.~\cite{Elgood:2020nls} two of us applied these ideas to the (by far)
  more complicated case of the heterotic superstring effective action to first
  order in $\alpha'$, obtaining the first law of black hole mechanics with
  work terms\footnote{Not all work terms were recovered in that article: at
    the time it was not known how to obtain the work terms proportional to the
    variations of the magnetic charges in this formalism (a problem solved in
    Ref.~\cite{Ortin:2022uxa}), nor the terms proportional to the asymptotic
    values of the scalars found in Ref.~\cite{Gibbons:1996af} (a problem
    solved in Ref.~\cite{Ballesteros:2023iqb}). Furthermore, there are charges
    associated to fields that only enter the theory after dimensional
    reduction, such as Kaluza-Klein vectors and it was not known how to
    recover the work terms proportional to the variations of those charges
    using the 10-dimensional action. This problem has been partially solved in
    Ref.~\cite{kn:G-FOZ}, but more work is needed to understand the
    higher-dimensional origin of all the work terms that occur in the 4- or
    5-dimensional first laws.} and a Lorentz frame-independent and
  gauge-invariant expression for the entropy.

  In the frame typically used to construct the solutions, the entropy formula
  found in Ref.~\cite{Elgood:2020nls} can be written in the same form as the
  one that follows from the Iyer-Wald prescription, except for one
  coefficient.

  As we have argued before, it would be wrong to decide which formula is right
  by just comparing the Wald entropies of the $\alpha'$-corrected stringy
  black holes obtained by applying them with the microscopic entropies. A
  internal consistency check of these entropy formulas is badly
  needed.\footnote{As we have stressed, the Iyer-Wald entropy formula is
    Lorentz frame-dependent, and this is enough to discard it. However, since
    the entropy formula of Ref.~\cite{Elgood:2020nls} gives a numerically very
    close result, it does not hurt to momentarily keep it in the game for the
    sake of the argument.}

  The property that the correct Wald entropy must satisfy is, precisely, that
  it satisfies the first law: $\partial S/\partial M = 1/T$. It would be
  enough to check which entropy satisfies this property, but this can only be
  done with $\alpha'$-corrected non-extremal stringy black holes. Finding an
  $\alpha'$-corrected non-extremal 4-dimensional black hole and the
  non-extremal versions of the 3-charge black holes to carry out this test was
  one of the main motivations behind Refs.~\cite{Cano:2019ycn} and
  \cite{Cano:2022tmn}, respectively. In both cases it was the entropy computed
  using the Lorentz frame- and gauge-independent formula of
  Ref.~\cite{Elgood:2020nls} that was shown to satisfy the first law.
 
\item Having computed by internally-consistent methods the Wald entropy of
  stringy black-hole solutions whose $\alpha'$ corrections have been reliably
  computed as well it is tempting to proceed immediately to compare that
  macroscopic entropy with the microscopic one. However, only the entropies of
  corresponding systems must be compared; for instance, the entropy of the
  (heterotic version of the) Strominger-Vafa black hole, which can be
  considered the strong-coupling limit of a set of solitonic 5-branes (S5) and
  fundamental strings (F1) with momentum flowing along them can be compared
  withe the entropy of the heterotic superstring quantized on such a
  background.

  At zeroth order in $\alpha'$ the identification of the branes corresponding
  to a given black-hole solution is straightforward: it follows directly from
  the calculation of the conserved charges of the solution. The definitions of
  those conserved charges and their calculation are unambiguous at that order
  and it is based on integrals of terms that satisfy Gauss laws and give the
  same results whether they are integrated over the horizon or at spatial
  infinity. The identification of the charges with numbers of branes is also
  unambiguous.

  At first order in $\alpha'$ the definitions of the conserved charges contain
  a large number of higher-order terms. All these terms are necessary to
  obtain integrands that satisfy Gauss laws but, since they are very hard to
  compute and they vanish asymptotically much faster than the zeroth-order
  terms, they are typically ignored. These integrands give conserved charges
  when integrated at spatial infinity but they give different results when
  integrated over the horizon because the terms that have been ignored do not
  vanish there. Thus, we have different notions of charge \cite{Marolf:2000cb}
  with different properties of transformation under duality and the problem of
  figuring out which charges correspond to the numbers of
  charges\footnote{This problem is aggravated by the $\alpha'$-dependent
    definitions that the fields theat descend from the 10-dimensional
    Kalb-Ramond 2-form suffer in the dimensional reduction
    \cite{Elgood:2020xwu}.} that occur in the microscopic entropy and study
  their conservation.\footnote{Incidentally, this discussion casts a shadow
    over the results obtained using near-horizon solutions since only the
    horizon charges can be computed with them.} This is a difficult and not
  yet solved problem that will be studied in more detail elsewhere
  \cite{kn:ORZ},\footnote{Nevertheless, see the discussion concerning the S5
    brane at the end of Section~\ref{sec-discussion}.} but the reader should
  be aware of its existence when comparing the macroscopic and microscopic
  entropies at first and higher orders in $\alpha'$ because their agreement
  depends on the choice of variables.
    
\end{enumerate}

It should be clear form this discussion that this is a research program still
under way and far from having been completed: the $\alpha'$ corrections of
many black-hole solutions (multi-center, rotating...) and their entropies have
not yet been computed, the identification of the branes that give rise to the
corresponding string background has to be clarified and, at some point in the
future, one would like to extend all these results to higher orders in
$\alpha'$ and in the string coupling constant.

\begin{figure}[ht!]
\begin{center}
\includegraphics[scale=0.9,trim=40 0 0 0,clip]{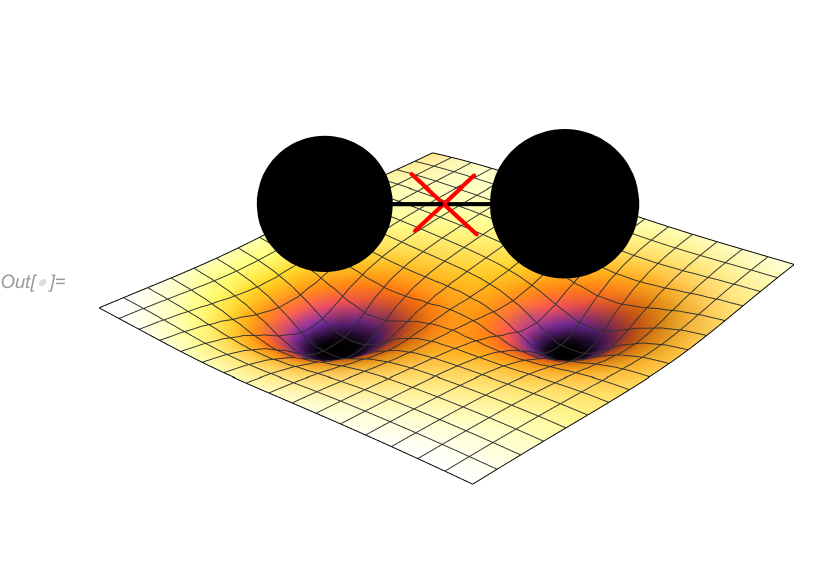}
\caption{The absence of struts (conical singularities) joining the centers is
  a necessary condition to interpret the solution as black holes in
  equilibrium without the help of external forces..}
\label{fig:nost1}
\end{center}
\end{figure}

\begin{figure}[ht!]
\begin{center}
\includegraphics[scale=0.9,trim=40 0 0 0,clip]{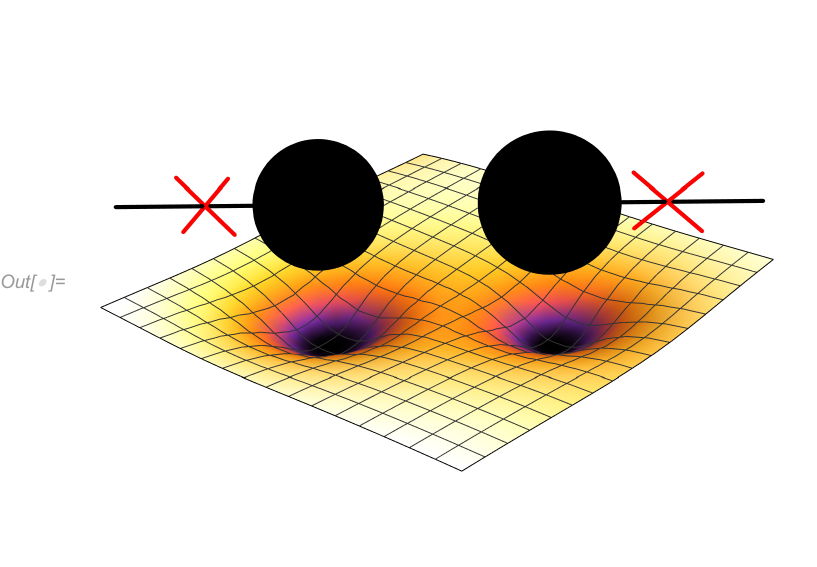}
\caption{The absance of struts (conical singularities) extending from the
  centers to infinity is another necessary condition to interpret the solution
  as black holes in equilibrium without the help of external forces..}
\label{fig:nost2}
\end{center}
\end{figure}

In this paper, as part of this program, we are going to focus on the $\alpha'$
corrections to the well-known multi-center generalizations of the 3- and
4-charge extremal, static, black-hole solutions in 5 and 4 dimensions. We are
interested in the corrections to the geometry and to the entropy and their
consequences. In particular, we want to know whether the $\alpha'$ corrections
preserve the regularity of the horizons and of the rest of the spacetime. It
has to be taken into account that solutions describing collinear Schwarzschild
black holes in static equilibrium have long been known
\cite{Israel-Kahn,Costa:2000kf}. These solutions, however, have conical
singularities in the line joining the centers or extending from the centers to
infinity, known as \textit{struts}, associated to the external forces
necessary to hold the system in equilibrium. The absence of these
singularities and the regularity of the horizons can be interpreted as a proof
of the cancellation of interaction energies and of the equilibrium of forces
between the black holes \cite{Brill:1963yv,Hartle:1972ya} (see
Figs.~\ref{fig:nost1} and \ref{fig:nost2}.\footnote{For an extended
  discussion on this we refer to the Introduction of \cite{Meessen:2017rwm}
  and references therein.}. We also want to know whether the fragmentation of
large black holes into smaller black holes is entropically favored.  We are
going to consider supersymmetric and some non-supersymmetric multi-center
black-hole solutions to see of and how the equilibrium of forces between the
black holes depends on supersymmetry.\footnote{In the solutions that we are
  going to consider, the existence of unbroken supersymmetry depends on the
  relative sign between two of the black-hole charges.  This relative sign
  does not cover all the non-supersymmetric possibilities in four dimensions,
  but the corrections to those not included here are much more difficult to
  deal with \cite{Cano:2021nzo}.}

Our calculation of the $\alpha'$ corrections in mainly based on
Ref.~\cite{Chimento:2018kop} in which a very general class of
$\alpha'$-corrected solutions was found. This class includes multi-center
black-hole solutions, as we are going to show.

This paper is organized as follows: in Section~\ref{sec-heteroticzeroth} we
review the five- and four-dimensional multi-center black hole solutions we
will work with at leading order in $\alpha'$, starting with their common
ten-dimensional origin. In Section~\ref{sec-heteroticfirst} we show the
$\alpha'$-corrected solutions and analyze their thermodynamic properties.
Finally, in Section~\ref{sec-discussion} we summarize our main results and
discuss how the $\alpha'$ corrections would affect the fragmentation of these
extremal black holes into smaller extremal black holes of the same kind. The
appendices contain the low-energy effective action and corresponding equations
of motion of the heterotic superstring to first order in $\alpha'$ and a
collection of results which are used in the main text.

\section{Heterotic multi-center black-hole solutions at leading order}
\label{sec-heteroticzeroth}

Let us begin with a review of two well-known families of multi-center
black-hole solutions in five and four dimensions which arise as solutions of
the heterotic superstring effective field theory compactified on a
torus.\footnote{This theory is described in Appendix~\ref{app-action}. At
  zeroth-order in $\alpha'$ and after compactification and truncation on
  T$^{5}$ one obtains a 5-dimensional supergravity whose supersymmetric
  solutions have been classified and studied in
  Ref.~\cite{Gauntlett:2002nw}. Some of them describe muticenter black-hole
  solutions.  Further compactification on a circle gives the 4-dimensional
  solutions we are going to consider. Their uplift to 10 dimensions gives the
  heterotic supersymmetric zeroth-order solutions we are describing in this
  section. In the 5-dimensional case, changing the global sign of the
  Kalb-Ramond field one trivially obtains a new solution, but breaks
  supersymmetry \cite{Cano:2021nzo}. In this way one obtains
  non-supersymmetric multi-center black-hole solutions. In the 4-dimensional
  case there are more global sign changes that transform solutions into
  solutions breaking supersymmetry \cite{Cano:2021nzo}. See also
  Ref.~\cite{Ortin:1996bz}. Particular cases of the 4-dimensional multi-center
  black hole solutions we are considering were found by Gibbons in
  Ref.~\cite{Gibbons:1982ih} and include the Majumdar-Papapetrou solutions
  Ref.s~\cite{kn:Ma,kn:Pa} (see also Ref.~\cite{Hartle:1972ya}). For a more
  recent account of multi-center black-hole solutions in String Theory see,
  \textit{e.g.}~\cite{VandenBleeken:2008tsa} and references therein.} Both
families have as a common origin a very general class of ten-dimensional
solutions which depend on a choice of four-dimensional hyper-K\"ahler metric
($m,n=1,2,3,4$),

\begin{equation}
  d\sigma^{2}
  =
  h_{\underline{m}\underline{n}}\,dx^{m}dx^{n}\,,
\end{equation}

\noindent
and on a choice of three functions,
$\mathcal{Z}_{0},\mathcal{Z}_{+},\mathcal{Z}_{-}$, which are harmonic in this
hyper-K\"ahler space, \textit{i.e.}~they satisfy the linear equations

\begin{equation}
  \label{eq:harmonicityconditions}
d\star_{\sigma}d\mathcal{Z}_{0,+,-}=0\,.
\end{equation}

\noindent
This linearity allows us to construct multi-center solutions by
``superposition'' of several single-center solutions.

In terms of these four objects, the string-frame line element $d\hat{s}^{2}$,
the Kalb-Ramond 3-form field strength $\hat{H}$ and the dilaton field
$\hat{\phi}$ of the solutions take the form\footnote{We use hats to
  distinguish the ten-dimensional fields from the five- and four-dimensional
  ones.}

\begin{subequations}
  \label{eq:10dsolution}
  \begin{align}
  d\hat{s}^{2}
  & = 
    \frac{1}{\mathcal{Z}_{+}\mathcal{Z}_{-}}dt^{2}
    -\mathcal{Z}_{0}\,d\sigma^{2}
    -\frac{k_{\infty}^{2}\mathcal{Z}_{+}}{\mathcal{Z}_{-}}
    \left[dz+k_{\infty}^{-1}
    \left(\mathcal{Z}^{-1}_{+}-1\right)dt\right]^{2}
    -d\vec{y}^{\,2}_{(4)}\,,
  \\
  & \nonumber \\
  \hat{H}
    & =
      d\left[\left(2\epsilon-1\right) k_{\infty}\left(\mathcal{Z}^{-1}_{-}-1\right)
 dt \wedge dz\right]
        +\star_{\sigma} d \mathcal{Z}_{0}\,,
  \\
    &  \nonumber \\
  e^{-2\hat{\phi}}
  & = 
       e^{-2\hat{\phi}_{\infty}}\mathcal{Z}_{-}/\mathcal{Z}_{0}\,.
  \end{align}
\end{subequations}

\noindent
The modulus $\hat{\phi}_{\infty}$ will correspond the asymptotic value of the
dilaton if, as we will assume, the functions $\mathcal{Z}_{-}$ and
$\mathcal{Z}_{0}$ asymptote to $1$. Since the vacuum expectation value of the
exponential of the dilaton is the string coupling constant, for
asymptotically-flat solutions we can make the identification,

\begin{equation}
\hat{g}_{s}=e^{{\hat\phi}_{\infty}}\,.
\end{equation}

\noindent
The coordinates $\vec{y}_{(4)}$, periodically identified with equal periods
$2\pi \ell_{s}$, where $\ell_{s}$ is the string length, parametrize a
four-dimensional torus $\mathbb{T}^{4}$ which has trivial internal dynamics,
meaning that all the Kaluza-Klein (KK) zero modes (scalars and vectors) are
trivial.\footnote{In particular, the KK scalars associated to the ${\mathbb T}^{4}$ are equal to equal to their vacuum expectation values that we set to $1$ for convenience, as they do not play any r\^ole in our solutions. } The coordinate $z \sim z+2\pi \ell_{s}$ further parametrizes a circle
$\mathbb{S}^{1}_{z}$ whose asymptotic radius $R_{z}$ is related to the
asymptotic value of the KK scalar associated to this compact direction,
$k_{\infty}$, and to $\ell_{s}$ by

\begin{equation}
R_{z} = k_{\infty} \ell_{s}\,.  
\end{equation}

\noindent
Finally, $\epsilon$ is the supersymmetry-breaking parameter. It takes two
values, $\epsilon=\{0, 1\}$, which correspond to the relative sign between two
of the black hole charges. It turns out that the configuration is
supersymmetric only if $\epsilon=1$ \cite{Cano:2021nzo}. We have not
considered the possibility of changing the sign of the second term in
$\hat{H}$. This sign becomes the sign of a third black-hole charge upon
dimensional reduction. At zeroth order, the effect of this change of sign is
almost trivial, but the first-order in $\alpha'$ corrections are much more
difficult to obtain than with the chosen sign \cite{Cano:2021nzo}.

Many different solutions can be obtained from Eq.~(\ref{eq:10dsolution}) by
making particular choices of hyper-K\"ahler space and harmonic functions. Many
of them are singular or have no known physical interpretation. Let us see
which choices lead to five- and four-dimensional multi-center black-hole
solutions.

\subsection{Multi-center black-hole solutions in five dimensions}

The simplest choice of hyper-K\"ahler space which allows for five-dimensional
multi-center black-hole solutions is $\mathbb{E}^{4}$. The choice of harmonic
functions in $\mathbb{E}^{4}$ that gives rise to multi-center three-charge
black holes is the following,\footnote{It has been recently proven in
  \cite{Lucietti:2020ryg} that, in the Einstein-Maxwell theory, the standard
  multi-center black-hole solutions are the only ones within the Majumdar-Papapetrou
  class that give rise to asymptotically-flat black holes with regular event
  horizons. This extends a previous result by Chrusciel and Tod
  \cite{Chrusciel:2006pc} to $d>4$ dimensions.}

\begin{equation}
  \label{eq:Zs5d}
  \mathcal{Z}_{+}
  =
  1+\sum_{a=1}^{n_{c}}\frac{q^{a}_{+}}{\rho^{2}_{a}}\,,
  \hspace{1cm}
  \mathcal{Z}_{-}
  =
  1+\sum_{a=1}^{n_{c}}\frac{q^{a}_{-}}{\rho^{2}_{a}}\,,
  \hspace{1cm}
  \mathcal{Z}_{0}=1+\sum_{a=1}^{n_{c}}\frac{q^{a}_{0}}{\rho^{2}_{a}}\,,
\end{equation}

\noindent
where $n_{c}$ is the total number of centers and

\begin{equation}
  \label{eq:rhoa}
\rho^{2}_{a}=\left(x^{m}-x^{m}_a\right)\left(x^{m}-x^{m}_a\right)\,,
\end{equation}

\noindent
$x^{m}_{a}$ being the position of the $a^{\rm th}$ center in
$\mathbb{E}^{4}$. The parameters $q^{a}_{+}, q^{a}_{-}$ and $q^{a}_{0}$ must
be taken to be strictly positive in order to avoid naked singularities and,
then, they correspond to the absolute value of the charges associated to each
black hole, up to normalization factors.

The five-dimensional form of the solution is obtained by compactification over
the internal manifold, $\mathbb{T}^{4}\times \mathbb{S}^{1}_{z}$. This
operation can be conveniently carried out in two steps. Firstly, we compactify
over the trivial $\mathbb{T}^{4}$. The corresponding six-dimensional solution
has exactly the same form as the ten-dimensional one except for the term
$-d\vec{y}^{\,2}_{(4)}$ in the metric, which is not present in the former.
Secondly, we reduce the six-dimensional solution over
$\mathbb{S}^{1}_{z}$. Using the relations between the higher- and
lower-dimensional fields provided in
Appendix~\ref{app:dimensionalreductiontorus}, we obtain

\begin{subequations}
  \begin{align}
    \label{eq:esametrica}
    ds^{2}_{{\rm E}}
    & = 
      \left(\mathcal{Z}_{+}\mathcal{Z}_{-}\mathcal{Z}_{0}\right)^{-2/3}dt^{2}
      -\left(\mathcal{Z}_{+}\mathcal{Z}_{-}\mathcal{Z}_{0}\right)^{1/3}dx^{m}dx^{m}\,,
    \\
& \nonumber \\
    H
    & =
      \tfrac{1}{3!}\epsilon_{mnpq}\partial_q \mathcal{Z}_{0}\,dx^{m}\wedge
      dx^{n}\wedge dx^{p}\,,
    \\
    & \nonumber \\
    F
    & =
      d\left[k_{\infty}^{-1}\left(\mathcal{Z}^{-1}_{+}-1\right)dt\right] \,,
    \\
& \nonumber \\
    G
    & =
     d\left[ \left(2\epsilon-1\right)k_{\infty}\left(\mathcal{Z}^{-1}_{-}-1\right)dt\right]\,,
    \\
& \nonumber \\
    e^{-2\phi}
    & =
      e^{-2\phi_{\infty}}\mathcal{Z}^{1/2}_{+}\mathcal{Z}^{1/2}_{-}\mathcal{Z}^{-1}_{0}\,,
    \\
    & \nonumber \\
    k
    & =
      k_{\infty} \mathcal{Z}^{1/2}_{+}\mathcal{Z}^{-1/2}_{-}\,,
  \end{align}
\end{subequations}

\noindent
where $ds^{2}_{\rm E}$ stands for the five-dimensional line element in the
so-called \emph{modified Einstein frame} \cite{Maldacena:1996ky},

\begin{equation}
F = dA\,,
\hspace{1cm}
G = dC\,,
\end{equation}

\noindent
are, respectively, the field strengths of the KK and winding vector fields, $A$ and $C$,

\begin{equation}
  H
  =
  dB-\tfrac{1}{2}A\wedge G -\tfrac{1}{2}C \wedge F\,,
\end{equation}

\noindent
is the 3-form field strength of the Kalb-Ramond 2-form $B$, $k$ is the KK
scalar associated to the internal direction $z$ and $\phi$ is the
five-dimensional dilaton. In five dimensions, the Kalb-Ramond field can be dualized
into a vector field whose field strength is

\begin{equation}
K = d\left[\left(\mathcal{Z}_{0}^{-1}-1\right)dt\right]\,.  
\end{equation}

\subsubsection{Thermodynamics}

The solution is asymptotically flat and its only potential singularities sit
at the centers $x^{m}=x^{m}_{a}$. If the charge products
$q^{a}_{+}q^{a}_{-}q^{a}_{0}$ are finite, though, these are just coordinate
singularities: near the $a^{\rm th}$ center, the metric is that of
AdS$_{2}\times \mathbb{S}^{3}$, which is one of the possible near-horizon
geometries of extremal black holes in five dimensions
\cite{Kunduri:2013gce}. Hence, the $a^{\rm th}$ center is, actually a
two-sphere of finite radius $(q^{a}_{+} q^{a}_{-}q^{a}_{0})^{1/6}$ which
corresponds to the event horizon of an extremal black hole. The
Bekenstein-Hawking (BH) entropy of the $a^{\rm th}$ black hole is, then,

\begin{equation}
  S^{a}_{\rm {BH}}
  =
  \frac{\pi^{2}}{2 G^{(5)}_{\rm N}}\sqrt{q^{a}_{+} q^{a}_{-}q^{a}_{0}}\,.
\end{equation}

\noindent
The entropy of the solution is just the sum of the entropies of all the black
holes:

\begin{equation}
  S_{\rm {BH}}
  =
  \sum_{a} S^{a}_{\rm {BH}}
  =
  \frac{\pi^{2}}{2 G^{(5)}_{\rm N}}\sum_{a}\sqrt{q^{a}_{+} q^{a}_{-}q^{a}_{0}}\,.
\end{equation}

Strictly speaking, only the total mass of the solution can be rigorously
defined. From the asymptotic expansion of the $tt$ component of the metric we
find that it is given by

\begin{equation}
  \label{eq:0thordermass}
  M
  =
  \frac{\pi}{4G^{(5)}_{\rm N}} \sum_{a}\left(q^{a}_{+} +q^{a}_{-}+q^{a}_{0}\right)\,.
\end{equation}

\noindent
However, since the vector fields of the solution are Abelian, we can compute
the charges of each black hole independently and, as we have already
mentioned, these are given given by the parameters
$q^{a}_{+}, q^{a}_{-},q^{a}_{0}$ up to signs and normalization. An isolated
extremal black hole with those charges would have a mass given by

\begin{equation}
M^{a} =  \frac{\pi}{4G^{(5)}_{\rm N}} \left(q^{a}_{+} +q^{a}_{-}+q^{a}_{0}\right)\,.
\end{equation}

\noindent
Since the total mass is the sum of the would-be individual masses of all the
black holes,

\begin{equation}
M = \sum_{a=1}^{n_{c}}M^{a}\,.  
\end{equation}

\noindent
This fact implies that the black-holes mutual interaction energies (all the
Einstein equations know about \cite{Meessen:2017rwm}) are zero because there is
a cancellation between all the different contributions (gravitational,
electromagnetic etc.) The cancellation of these interaction energies is
associated to a cancellation of the forces between the centers and, then,
since the solution is static, it is reasonable to conclude that the solution
contains $n_{c}$ extremal black holes with those charges and masses in
equilibrium. The absence of singularities indicates that no external forces
are necessary to keep those black holes in equilibrium.

\subsubsection{Microscopic description}

The string-theory description of these solutions is well known: they can be
understood as a superposition of solitonic 5-branes (S5) ---often referred to as NS5 branes---wrapped around the directions parametrized by the coordinates $y^{1},\cdots,y^{4},z$, fundamental strings (F1) wound around the circle parametrized by $z$ and waves (W)
carrying momentum propagating along the same circle, as we illustrate in Table~\ref{diagram5d}.

\begin{table}[ht!]
\begin{center}
\begin{tabular}{ccccccccccc}
&$t$&$z$&$y^{1}$&$y^{2}$&$y^{3}$&$y^{4}$&$x^{1}$&$x^{2}$&$x^{3}$&$x^{4}$\\
\hline
F1&$\times$ &$\times$&$\sim$&$\sim$&$\sim$&$\sim$&$-$&$-$&$-$&$-$\\
\hline
W&$\times$&$\times$&$\sim$&$\sim$&$\sim$&$\sim$&$-$&$-$&$-$&$-$\\
\hline
S5&$\times$&$\times$&$\times$&$\times$&$\times$&$\times$&$-$&$-$&$-$&$-$\\
\end{tabular}
\caption{Sources associated to five-dimensional black holes. The symbol
  $\times$ stands for the worldvolume directions and $-$ for the transverse
  directions. The symbol $\sim$, in turn, denotes a transverse direction over
  which the corresponding object has been smeared.  }
\label{diagram5d}
\end{center}
\end{table}

From the lower-dimensional point of view, these fundamental objects act as
point-like sources for the three different types of charges carried by our
solutions. The locations of the sources in the non-compact space coincides
with the position of the centers, where the Laplace
Eqs.~(\ref{eq:harmonicityconditions}) are not satisfied unless we take into
account the coupling of the sources to the background fields. This give rise
to a sum of Dirac delta functions, $\delta^{(4)}(x-x_{a})$,\footnote{Strictly
  speaking, this is just an effective description of the actual situation: the
  centers $x^{m}=x^{m}_{a}$ are not points, but finite-radii two-spheres from
  which the Abelian fields' fluxes ``emerge'' in radial direction.} each of
which is weighted by the fraction of the total tension/charge associated to
all the sources located at $x^{m}=x^{m}_{a}$, which in turn is proportional to
the amounts of momentum, winding and S5 branes in each of the centers,
respectively denoted by $n^{a}$, $w^{a}$ and $N^{a}$.\footnote{These
  parameters are assumed to be positive.}  This is exactly the same kind of
contribution that one gets from the Laplacian of $\mathcal{Z}_{+,-,0}$, with
the only difference that now the delta functions are weighted by
$q^{a}_{+}, q^{a}_{-}$ and $q^{a}_{0}$. Imposing that both contributions
cancel (\textit{i.e.}~that the equations of motion are also satisfied at the
centers), one finds the following relation between both sets of parameters
\cite{Cano:2018qev}:

\begin{equation}\label{eq:sources5d}
  q^{a}_{+}=\frac{g_{s}^{2}{\alpha'}^{2}}{R_{z}^{2}}\,n^{a}\,,
  \hspace{1cm}
  q^{a}_{-}=g_{s}^{2}{\alpha'}\,w^{a}\,,
  \hspace{1cm}
q^{a}_{0}=\alpha'\,N^{a}\,.
\end{equation}

Using these relations and the expression of the five-dimensional Newton
constant in terms of stringy variables

\begin{equation}
G^{(5)}_{\rm N}=\frac{\pi g_{s}^{2} {\alpha'}^{2}}{4R_{z}}\,,
\end{equation}

\noindent
one can rewrite the Bekenstein-Hawking entropy and the mass in terms of the
source parameters as follows

\begin{subequations}
  \begin{align}
  S_{\rm {BH}}
  & = 2\pi \sum_{a}\sqrt{n^{a} w^{a} N^{a}}\,,
  \\
  & \nonumber \\
  M
  & =
    \frac{n}{R_{z}}+\frac{R_{z}}{\alpha'}\, w
    +\frac{R_{z}}{g_{s}^{2}\alpha'}\, N\,,
  \end{align}
\end{subequations}

\noindent
where $n=\sum_{a} n^{a}$, $w=\sum_{a} w^{a}$ and $N=\sum_{a} N^{a}$ are the
total momentum, winding and number of S5-branes respectively. As we can see,
the total energy of the configuration only depends on the absolute value of
the total charges, no matter how they are distributed among the different
centers. This will no longer be the case in presence of higher-derivative
corrections, as we will see Section~\ref{sec-heteroticfirst}.

\subsection{Multi-Center black-hole solutions in four dimensions}

In order to describe multi-center black-hole solutions in four dimensions, one of the
directions of the hyper-K\"ahler space must be compactified, which in practice
means that it has to be isometric. Supersymmetry is preserved in the
compactification along this isometric direction if the isometry is consistent
with the hyper-K\"ahler structure, \textit{i.e.}~if the isometry is
\textit{triholomorphic}. In this case, the hyper-K\"ahler space must belong to
the Gibbons-Hawking class\cite{Gibbons:1979zt,Gibbons:1987sp}.

As we will see in the next section, there is another advantage for considering
Gibbons-Hawking spaces: their contribution to the $\alpha'$-corrected Bianchi
identity takes a form (the Laplacian of a function) that allows us to solve
the Bianchi identity in a fully analytic fashion \cite{Chimento:2018kop}.

If $w\sim w +2\pi \ell_{s}$ is the compact coordinate adapted to the isometry,
a Gibbons-Hawking metric can be written in the form

\begin{equation}
  d\sigma^{2}
  =
  h_{\underline{m}\underline{n}}\,dx^{m} dx^{n}
  =
  \ell^{2}_{\infty}\mathcal{H}^{-1}\left(dw+\ell^{-1}_{\infty}\chi\right)^{2}
  +\mathcal{H}\,d\vec{x}^{\,2}_{(3)}\,,
\end{equation}

\noindent
where $\mathcal{H}$ is the Gibbons-Hawking function and $\chi$ is a 1-form
which satisfying the equation

\begin{equation}
d\chi=\star_{(3)}\,d\mathcal{H}\,,
\end{equation}

\noindent
where $\star_{(3)}$ the Hodge star operator in $\mathbb{E}^{3}$. The above
equation implies that $\mathcal{H}$ is harmonic in $\mathbb{E}^{3}$. This
property is also satisfied by the functions $\mathcal{Z}_{+,-,0}$ if they are
harmonic in $\mathbb{E}^{4}$ and we demand that they do not depend on the
isometric coordinate $w$. The modulus $\ell_{\infty}$ corresponds to the
asymptotic value of the KK scalar $\ell$ associated to the circle parametrized
by $w$. The asymptotic radius of this circle is given in terms of
$\ell_{\infty}$ an the string length $\ell_{s}$ by

\begin{equation}
R_{w}=\ell_{\infty}\ell_{s}\,.
\end{equation}

The choice that allows us to describe multi-center black-hole solutions in
four dimensions is

\begin{equation}
  \label{eq:Zs4d}
\mathcal{Z}_{+}=1+\sum_{a=1}^{n_{c}}\frac{q^{a}_{+}}{r_{a}}\,,
\hspace{0.5cm}
\mathcal{Z}_{-}=1+\sum_{a=1}^{n_{c}}\frac{q^{a}_{-}}{r_{a}}\,,
\hspace{0.5cm}
\mathcal{Z}_{0}=1+\sum_{a=1}^{n_{c}}\frac{q^{a}_{0}}{r_{a}}\,,
\hspace{0.5cm}
\mathcal{H}=1+\sum_{a=1}^{n_{c}}\frac{q^{a}_{\mathcal{H}}}{r_{a}}\,,
\end{equation}

\noindent
where $r_{a}=||\vec{x}-\vec{x}_{a}||$. As before, $\vec{x}_{a}$ denotes the
position of the $a^{\rm th}$ center in $\mathbb{E}^{3}$ and the parameters
$q^{a}_{+}, q^{a}_{-}, q^{a}_{0}$ and $q^{a}_{\cal H}$ are assumed to be
strictly positive.

The 1-form $\chi$ is given, locally, by

\begin{equation}
  \chi
  =
  \sum_{a} q^{a}_{\mathcal{H}} \cos{\theta_{a}} d\phi_{a}\,,
\end{equation}

\noindent
where $r_{a},\theta_{a},\phi_{a}$ are spherical coordinates associated to the
$a^{\rm th}$ center

\begin{equation}
  \vec{x}-\vec{x}_{a}
  =
  (r_{a} \sin{\theta_{a}} \cos{\phi_{a}}, r_{a} \sin{\theta_{a}}
  \sin{\phi_{a}}, r_{a} \cos{\theta_{a}})\,.
\end{equation}

With these choices, the ten-dimensional solution can be rewritten locally in
the form

\begin{subequations}
  \label{eq:10dsolution-2}
  \begin{align}
    d\hat{s}^{2}
    & =
          \frac{1}{\mathcal{Z}_{+}\mathcal{Z}_{-}}dt^{2}
      -\ell_{\infty}^{2}\frac{\mathcal{Z}_{0}}{\mathcal{H}}
      \left(dw+\ell_{\infty}^{-1}\chi\right)^{2}
-\mathcal{Z}_{0}\mathcal{H}d\vec{x}_{(3)}^{\,2}
    \nonumber \\
  & \nonumber \\
  & 
    -k_{\infty}^{2}\frac{\mathcal{Z}_{+}}{\mathcal{Z}_{-}}
    \left[dz +k_{\infty}^{-1}
    \left(\mathcal{Z}^{-1}_{+}-1\right)dt\right]^{2}
      -d\vec{y}^{\,2}_{(4)}\,,
\\
&  \nonumber \\
\hat{H}
    & =
      d\left[\left(2\epsilon-1\right)
      k_{\infty}\left(\mathcal{Z}^{-1}_{-}-1\right) dt \wedge dz\right]
      +\ell_{\infty}\star_{(3)}d\mathcal{Z}_{0}\, \wedge dw \,,
      \\
& \nonumber \\
e^{-2\hat{\phi}}
& = 
e^{-2\hat{\phi}_{\infty}}
\mathcal{Z}_{-}/\mathcal{Z}_{0}\,.
\end{align}
\end{subequations}

In order to avoid Dirac-Misner string singularities we have to impose the
following ``quantization conditions'' on the $q^{a}_{\mathcal{H}}$s':

\begin{equation}
  \label{eq:KKquantization}
  q^{a}_{\mathcal{H}}=\frac{R_{w} W^{a}}{2}\,,
  \hspace{1cm}
  W^{a} \in \mathbb{Z}^{+}\,\,\,\,\forall a\,.
\end{equation}

The four-dimensional solution is obtained by compactifying the above solution
on the six-torus
$\mathbb{T}^{6}=\mathbb{T}^{4}\times \mathbb{S}^{1}_{z}\times
\mathbb{S}^{1}_{w}$ by using (twice) the results in
Appendix~\ref{app:dimensionalreductiontorus}. It has the following
non-vanishing fields,

\begin{subequations}
  \begin{align}
    ds^{2}_{\rm E}
    & =
\left(\mathcal{Z}_{+}\mathcal{Z}_{-}\mathcal{Z}_{0}\mathcal{H}\right)^{-1/2}dt^{2}
 -\left(\mathcal{Z}_{+}\mathcal{Z}_{-}\mathcal{Z}_{0}\mathcal{H}\right)^{1/2}
  d\vec{x}^{\,2}_{(3)}\,,
    \\
   & \nonumber \\
   \left(A^{\alpha}\right)
   & =
   \left(
     \begin{array}{c}
\ell_{\infty}^{-1}\chi  \\       
       \\
       k^{-1}_{\infty}\left(\mathcal{Z}^{-1}_{+}-1\right)dt \\
     \end{array}
    \right)\,,
    \\
  &  \nonumber \\
  \left(C_{\alpha}\right)
  & =
  \left(
    \left(2\epsilon-1\right)k_{\infty}\left(\mathcal{Z}^{-1}_{-}-1\right)  dt\,,
    \,\,\,\,
      \ell_{\infty}\chi_{0}
    \right)\,,
    \\
    & \nonumber \\
      \left(G_{\alpha \beta}\right)
  & =
  \left(
    \begin{array}{cc}
     \ell_{\infty}^{2}\mathcal{Z}_{0}/\mathcal{H}  & 0 \\
                                                            & \\
      0 & k_{\infty}^{2}\mathcal{Z}_{+}/\mathcal{Z}_{-} \\
     \end{array}
   \right)\,,
    \\
    & \nonumber \\
      e^{-2\phi}
    & =
      e^{-2\phi_{\infty}}
  \sqrt{\frac{\mathcal{Z}_{+}\mathcal{Z}_{-}}{\mathcal{Z}_{0}\mathcal{H}}}\,,
  \end{align}
\end{subequations}

\noindent
where

\begin{equation}
e^{-2\phi_{\infty}} =  e^{-2\hat{\phi}_{\infty}}k_{\infty}\ell_{\infty}\,,
\end{equation}
and 

\begin{equation}
\chi_{0}=\sum_{a} q^{a}_{0}\cos \theta_{a} d\phi_{a}\,.
\end{equation}

\subsubsection{Thermodynamics}

The reasoning leading to the identification of the locations of the poles of
the harmonic functions with extremal black-hole event horizons is identical to
that of the five-dimensional case. Now, the near-horizon geometries are
AdS$_{2}\times \mathbb{S}^{2}$ and the radius of the $a^{\rm th}$ of these
spaces is $(q^{a}_{+}q^{a}_{-}q^{a}_{0}q^{a}_{\mathcal{H}})^{1/4}$. The
Bekenstein-Hawking entropy to the $a^{\rm th}$ black hole is

\begin{equation}
  \mathrm{S}^{a}_{\rm BH}
  =
  \frac{\pi}{G_{N}^{(4)}}\sqrt{q^{a}_{+}q^{a}_{-}q^{a}_{0}q^{a}_{\mathcal{H}}}\,,
\end{equation}

\noindent
and the entropy of the complete spacetime is its sum

\begin{equation}
  S_{\rm{BH}}
  =
  \sum_{a}  \mathrm{S}^{a}_{\rm BH}\,.
\end{equation}

We can also assign a mass $M^{a}$ to the $a^{\rm th}$ center

\begin{equation}
  M^{a}
  =
  \frac{1}{4G^{(4)}_{\rm N}}
  \left(q^{a}_{+}+q^{a}_{-}+q^{a}_{0}+q^{a}_{\mathcal{H}}\right)\,,
\end{equation}

\noindent
and the cancellation of the mutual interaction energies follows from the
relation between the total mass $M$ and the putative masses of the individual
black holes

\begin{equation}
  M
  =
  \sum_{a}M^{a}
  =
  \frac{1}{4G^{(4)}_{\rm N}} \left(\sum_{a}q^{a}_{+}
    +\sum_{a}q^{a}_{-}+\sum_{a}q^{a}_{0}+\sum_{a}q^{a}_{\mathcal{H}}\right)\,.
\end{equation}

\subsubsection{Microscopic description}

The main difference between the four- and five-dimensional cases is that in
the four-dimensional case there is one more type of charge: the magnetic
charge associated to the Kaluza-Klein vector. The sources for these charges
are Kaluza-Klein monopoles. Their topological charges are the integers
$W^{a}$, whose relation with the magnetic charges $q^{a}_{\mathcal{H}}$ was
found in Eq.~(\ref{eq:KKquantization}) by imposing the absence of Dirac-Misner
string singularities. The remaining sources (fundamental strings, waves and S5-branes) are those of the five-dimensional black holes smeared over the
coordinate $w$ which is transverse to their worldvolume directions, see
Table~\ref{diagram4d}.

\begin{table}[ht!]
\begin{center}
\begin{tabular}{ccccccccccc}
&$t$&$z$&$y^{1}$&$y^{2}$&$y^{3}$&$y^{4}$&$w$&$x^{1}$&$x^{2}$&$x^{3}$\\
\hline
F1&$\times$ &$\times$&$\sim$&$\sim$&$\sim$&$\sim$&$\sim$&$-$&$-$&$-$\\
\hline
W&$\times$&$\times$&$\sim$&$\sim$&$\sim$&$\sim$&$\sim$&$-$&$-$&$-$\\
\hline
S5&$\times$&$\times$&$\times$&$\times$&$\times$&$\times$&$\sim$&$-$&$-$&$-$\\
\hline
KK6&$\times$&$\times$&$\times$&$\times$&$\times$&$\times$&$\sim$&$-$&$-$&$-$\\
\end{tabular}
\caption{Sources associated to the four-dimensional black holes. }
\label{diagram4d}
\end{center}
\end{table}

Due to this smearing, the relations between the parameters $q^{a}$ that occur
in the harmonic functions and the source parameters are no longer given by
Eqs.~(\ref{eq:sources5d}), but instead by
\cite{Cano:2018brq}\footnote{Although we use the same symbols for the
  parameters that occur in the harmonic functions for the five- and
  four-dimensional cases, they are different. In particular, they have
  different dimensions.}

\begin{equation}
  \label{eq:sources4d}
  q^{a}_{+}
  =
  \frac{g_{s}^{2}{\alpha'}^{2}}{2R_{z}^{2}R_{w}}\,n^{a}\,,
  \hspace{1cm}
  q^{a}_{-}
  =
  \frac{g_{s}^{2}{\alpha'}}{2R_{w}}\,w^{a}\,,
  \hspace{1cm}
  q^{a}_{0}
  =
  \frac{\alpha'}{2R_{w}}\,N^{a}\,.
\end{equation}

In terms of the source parameters, $n^{a},w^{a},N^{a}$ and $W^{a}$, the
entropies and putative masses of each of the centers read

\begin{subequations}
  \begin{align}
    S_{\rm BH}^{a}
    & =
      2\pi \sqrt{n^{a} w^{a} N^{a}W^{a}}\,,
    \\
    & \nonumber \\
    M^{a} & =
            \frac{n^{a}}{R_{z}}+\frac{R_{z}}{\alpha'}\, w^{a}
            + \frac{R_{z}}{g_{s}^{2}\alpha'}\, N^{a}
            +\frac{R^{2}_{w} R_{z}}{g_{s}^{2} {\alpha'}^{2}}W^{a}\,.
  \end{align}
\end{subequations}

The total entropy and mass are just the sums.

\section{Heterotic multi-center black-hole solutions at first order in $\alpha'$}
\label{sec-heteroticfirst}

It was shown in Ref.~\cite{Chimento:2018kop} that the first-order $\alpha'$
corrections to the general ten-dimensional heterotic background considered in
the previous section and given in Eqs.~(\ref{eq:10dsolution}) can be
analytically obtained in the supersymmetric case for arbitrary choices of
harmonic functions as long as the hyper-K\"ahler metric is a Gibbons-Hawking
space or $\mathbb{E}^{4}$. Here we extend the results of
\cite{Chimento:2018kop} to the $\epsilon=0$ case, which does not preserve any
supersymmetry.  In this section we will just describe the corrected
solutions.\footnote{Additional details are provided in
  Appendix~\ref{app:thesolution}.} They have exactly the same form as the
leading-order solutions described in Section~\ref{sec-heteroticzeroth} with
the only difference that the functions $\mathcal{Z}_{+}$ and $\mathcal{Z}_{0}$
get $\alpha'$ corrections.\footnote{This is only true for the 10-dimensional
  solutions, because the definitions of some the lower-dimensional fields (all
  those descending from the 10-dimensional Kalb-Ramond 2-form) have $\alpha'$
  corrections \cite{Elgood:2020xwu}. These will not be needed for our
  purposes: we only need the lower-dimensional metrics, whose expression in
  terms of the $\mathcal{Z}$ functions do not change.} For any solution to the
zeroth-order equations of motion of the form Eq.~(\ref{eq:10dsolution}), the
corrections to these functions are given by\footnote{Here we are using flat
  indices in the hyper-K\"ahler space $m,n,\ldots$, whose metric does not get
  any corrections.  Thus,
  $\partial_{m}X\partial_{m}Y \equiv
  h^{\underline{m}\underline{n}}\partial_{\underline{m}}X\partial_{\underline{n}}Y$.}

\begin{subequations}
  \begin{align}
    \label{eq:Zpcorrected}
    \mathcal{Z}_{+}
    & =
      \mathcal{Z}^{(0)}_{+}- \alpha'\left\{\frac{\epsilon}{2}
      \frac{\partial_{m}\mathcal{Z}^{(0)}_{+}\partial_{m}\mathcal{Z}^{(0)}_{-}}{\mathcal{Z}_{0}^{(0)}\mathcal{Z}_{-}^{(0)}}
      +\mathcal{H}_{+}\right\} + \mathcal{O}(\alpha^{\prime\, 2})\,,
    \\
    & \nonumber \\
    \label{eq:Z0corrected}
    \mathcal{Z}_{0}
    & =
      \mathcal{Z}_{0}^{(0)}+\alpha'\left\{\frac{1}{4}
      \left[\frac{\partial_{m}\mathcal{Z}_{0}^{(0)}\partial_{m}\mathcal{Z}_{0}^{(0)}}{\left(\mathcal{Z}_{0}^{(0)}\right)^{2}}
      +\frac{\partial_{m}\mathcal{H}^{(0)}\partial_{m}\mathcal{H}^{(0)}}{\left(\mathcal{H}^{(0)}\right)^{2}}\right]
      +\mathcal{H}_{0}\right\} +\mathcal{O}(\alpha^{\prime\, 2})\,,
  \end{align}
\end{subequations}

\noindent
where now $\mathcal{Z}^{(0)}_{+,-, 0}$ and $\mathcal{H}^{(0)}$ are the
zeroth-order value of those functions (\textit{i.e.}~harmonic functions in the
hyper-K\"ahler metric $h_{\underline{m}\underline{n}}$) and $\mathcal{H}_{0}$
and $\mathcal{H}_{+}$ are arbitrary harmonic functions in the hyper-K\"ahler
metric $h_{\underline{m}\underline{n}}$ (which does not get any corrections)
which will be determined later on by imposing appropriate boundary conditions.

The functions $\mathcal{H}$ and $\mathcal{Z}_{-}$ do not get any corrections,
hence they are simply given by

\begin{subequations}
  \begin{align}
    \mathcal{Z}_{-}
    & =
      \mathcal{Z}_{-}^{(0)}+\mathcal{O}(\alpha^{\prime\, 2})\,,
    \\
    & \nonumber \\
    \mathcal{H}
    & =
      \mathcal{H}^{(0)}+\mathcal{O}(\alpha^{\prime\, 2})\,.
  \end{align}
\end{subequations}

Let us now discuss the corrections to the multi-center black-hole solutions
studied in the previous section. 

\subsection{Multi-Center black-hole solutions in five dimensions}

Plugging the functions $\mathcal{Z}^{(0)}_{0}$ and $\mathcal{Z}^{(0)}_{+}$
chosen in Eqs.~(\ref{eq:Zs5d}) in Eqs.~(\ref{eq:Zpcorrected}) and
(\ref{eq:Z0corrected}), we get 

\begin{subequations}
  \begin{align}
  \mathcal{Z}_{+}
  & = 
    \mathcal{Z}_{+}^{(0)}
    +\alpha'\left[
    -2\epsilon\mathcal{Z}_{-}^{-1}\mathcal{Z}_{0}^{(0)\,-1}
    \sum_{a,b}\frac{q^{a}_{+}q^{b}_{-}n^{m}_{a}n^{m}_{b}}{\rho^{3}_{a}\rho^{3}_{b}}
    +\mathcal{H}_{+}\right]+\mathcal{O}(\alpha^{\prime\, 2})\,,
  \\
  & \nonumber \\
  \mathcal{Z}_{0}
  & =
    \mathcal{Z}_{0}^{(0)}
    +\alpha'\left[
    \mathcal{Z}_{0}^{(0)\,-2}\sum_{a,b}\frac{q^{a}_{0}q^{b}_{0}n^{m}_{a}n^{m}_{b}}{\rho^{3}_{a}\rho^{3}_{b}}
    +\mathcal{H}_{0}\right]+\mathcal{O}(\alpha^{\prime\, 2})\,,
  \end{align}
\end{subequations}

\noindent
where we have defined the unit radial vectors

\begin{equation}
n^{m}_{a}\equiv (x^{m}-x^{m}_{a})/\rho_{a}\,.
\end{equation}

We just have to determine the harmonic functions $\mathcal{H}_{+}$ and
$\mathcal{H}_{0}$.  We shall impose the following two conditions:

\begin{enumerate}
\item The solutions will be asymptotically flat with the following
  normalization of the functions: $\lim_{||x||\to
    \infty}\mathcal{Z}_{+,0}=1$. This implies

\begin{equation}
  \lim_{||x||\to \infty}\mathcal{H}_{+,0}=0\,.
\end{equation}

\item The coefficients of the $1/\rho_{a}^{2}$ poles of $\mathcal{Z}_{0,+}$ which
  arise in the $\rho_{a}\rightarrow 0$ will not be renormalized. That is

\begin{equation}
  \label{eq:boundaryconds}
  \left.\mathcal{Z}_{+}\right|_{x\to x_{a}}
  \sim
  \frac{q^{a}_{+}}{\rho^{2}_{a}}\,,
  \hspace{1cm}
  \left.\mathcal{Z}_{0}\right|_{x\to x_{a}}
  \sim
\frac{q^{a}_{0}}{\rho^{2}_{a}}\,.
\end{equation}
\end{enumerate}

\noindent
$\mathcal{H}_{0}$ and $\mathcal{H}_{+}$ must be  harmonic functions of the
same type as $\mathcal{Z}^{(0)}_{0}$ and $\mathcal{Z}^{(0)}_{+}$ to preserve
asymptotic flatness:

\begin{equation}
  \mathcal{H}_{0, +}
  =
  \alpha_{0, +}+ \sum_{a} \frac{\beta^{a}_{0,+}}{\rho^{2}_{a}}\,.
\end{equation}

\noindent
Then, the above conditions determine the coefficients
$\alpha_{0, +}, \beta^{a}_{0,+}$ as follows,

\begin{equation}
  \alpha_{0}=\alpha_{+}=0\,,
  \hspace{1cm}
  \beta^{a}_{0}=-1\,,
  \hspace{1cm}
  \beta^{a}_{+}=2\epsilon \frac{ q^{a}_{+}}{q^{a}_{0}}\,,
\end{equation}

\noindent
yielding

\begin{equation}
  \mathcal{H}_{+}
  =
  2\epsilon\sum_{a} \frac{ q^{a}_{+}}{q^{a}_{0}\rho^{2}_{a}}\,,
  \hspace{1cm}
\mathcal{H}_{0}
  =
  -\sum_{a}\frac{1}{\rho^{2}_{a}}\,.
\end{equation}

\noindent
Hence, the final form of the functions that get $\alpha'$ corrections is

\begin{subequations}
  \begin{align}
  \mathcal{Z}_{+}
  & = 
    \mathcal{Z}_{+}^{(0)}
    -2\epsilon\alpha'\left[
    \mathcal{Z}_{-}^{-1}\mathcal{Z}_{0}^{(0)\,-1}
    \sum_{a,b}\frac{q^{a}_{+}q^{b}_{-}n^{m}_{a}n^{m}_{b}}{\rho^{3}_{a}\rho^{3}_{b}}
    -\sum_{a}
\frac{q^{a}_{+}}{q^{a}_{0}\rho^{2}_{a}}\right]+\mathcal{O}(\alpha^{\prime\, 2})\,,
  \\
  & \nonumber \\
  \mathcal{Z}_{0}
  & =
    \mathcal{Z}_{0}^{(0)}
    +\alpha'\left[
    \mathcal{Z}_{0}^{(0)\,-1}
    \sum_{a,b}\frac{q^{a}_{0}q^{b}_{0}n^{m}_{a}n^{m}_{b}}{\rho^{3}_{a}\rho^{3}_{b}}
    -\sum_{a}
\frac{1}{\rho^{2}_{a}}\right]+\mathcal{O}(\alpha^{\prime\, 2})\,.
  \end{align}
\end{subequations}

The two conditions imposed to determine $\mathcal{H}_{0,+}$ have a physical
motivation. The first condition is equivalent to asking the
non-renormalization of the asymptotic value of the string coupling, the radius
of the internal direction and, therefore, of the five-dimensional Newton
constant. These constants defined the vacuum and, therefore, we are dealing
with corrected solutions in the original vacuum. This allows us to compare the
masses and the charges of the black holes before and after the corrections.

The second condition is meant to keep unmodified the relation between the
parameters of the solution $q^{a}_{0, +, -}$ and the physical parameters
specifying the microscopic system, $n^{a}, w^{a}$ and $N^{a}$
Eq.~\eqref{eq:sources5d}. As we have already discussed, this is due to the
fact that $n^{a}, w^{a}$ and $N^{a}$ are proportional to the coefficients of
the poles of the functions \cite{Cano:2018qev,Cano:2018brq}.  It is worth
remarking that this condition implies that the $1/\rho^{2}$ coefficient in the
asymptotic expansion of the functions can be corrected. Indeed, we find

\begin{equation}
  \mathcal{Z}_{0}
  \sim
  1+ \frac{1}{\rho^{2}}\left(\sum_{a} q^{a}_{0} -\alpha' n_{c}\right)+\dots\,,
\hspace{0.5cm}
\mathcal{Z}_{+}
\sim
1+ \frac{1}{\rho^{2}}
\sum_{a} \left(q^{a}_{+}+ \frac{2\epsilon\alpha' q^{a}_{+}}{q_{0}^{a}}\right)+\dots\,.
\end{equation}

Another natural choice of harmonic functions $\mathcal{H}_{0,+}$ would have
been to simply set both of them to zero. In such case, the $1/\rho^{2}$
coefficients in the asymptotic expansion would remain invariant, while the
coefficients of the poles of the functions would be corrected, contrarily to
what happens with the choice we have made. Nevertheless, it is important to
emphasize that both choices are physically equivalent since they give rise to
the same solution parametrized in very different ways. For instance, the
identifications in Eq.~\eqref{eq:sources5d} are no longer valid if we choose
$\mathcal{H}_{0,+}=0$.

In any case, these ambiguities disappear once the solution is expressed in
terms of the physical quantities. The behavior of
the functions near the $a^{\rm th}$ center is given by

\begin{equation}\label{eq:nhbehavior}
  \mathcal{Z}_{+}\sim\frac{g_{s}^{2}\alpha^{\prime\, 2} n^{a}}{R^{2}_{z}\rho^{2}_{a}}\,,
  \hspace{1cm}
\mathcal{Z}_{-}\sim\frac{g_{s}^{2}\alpha' w^{a}}{\rho^{2}_{a}}\,,
\hspace{1cm}
\mathcal{Z}_{0}\sim\frac{\alpha' N^{a}}{\rho^{2}_{a}}\,,
\end{equation}

\noindent
while, asymptotically, we find 

\begin{equation}
\mathcal{Z}_{+}\sim\,1+\frac{g_{s}^{2}\alpha^{\prime\, 2} |\mathcal{Q}_{+}|}{R^{2}_{z}\rho^{2}}\,,
\hspace{1cm} 
\mathcal{Z}_{-}\sim\,1+ \frac{g_{s}^{2}\alpha' |\mathcal{Q}_{-}|}{\rho^{2}}\,,
 \hspace{1cm}
\mathcal{Z}_{0}\sim\,1+ \frac{\alpha' |\mathcal{Q}_{0}|}{\rho^{2}}\,,
\end{equation}

\noindent
where we have defined, for later convenience,

\begin{equation}\label{eq:charges5d}
  \mathcal{Q}_{+}
  =
  n+2\epsilon\sum_{a}\frac{n^{a}}{N^{a}} \,,
  \hspace{1.5cm}
  \mathcal{Q}_{-}
  =
  \left(2\epsilon-1\right)w\,,
  \hspace{1.5cm}
  \mathcal{Q}_{0}
  =
  -\left(N-n_{c}\right)\,,
\end{equation}

\noindent
which coincide with the asymptotic charges of the solution, defined as

\begin{subequations}
\begin{align}
  {\ell}^{-1}_{s}\mathcal{Q}_{+}
  & =
    \frac{{g}_{s}^{2}}{16\pi G^{(5)}_{\rm
    N}}\int_{\mathbb{S}^{3}_{\infty}}e^{-2 \phi}k^{2} \;
    {\star} \,{ F}\,,
  \\
  & \nonumber \\
  T_{F1}\mathcal{Q}_{-}
  & =
    \frac{\hat{g}_{s}^{2}}{16\pi G^{(10)}_{\rm
    N}}\int_{\mathbb{T}^{4}\times\mathbb{S}^{3}_{\infty}}
    e^{-2\hat{\phi}} \;\hat{\star} \,\hat{H}\,,
  \\
    & \nonumber \\
  T_{S5}\mathcal{Q}_{0}
  & =
    \frac{1}{16\pi G^{(10)}_{\rm N}}\int_{\mathbb{S}^{3}_{\infty}}\hat{H}\,,
\end{align}
\end{subequations}

\noindent
where

\begin{equation}
T_{F1}=\frac{1}{2\pi\alpha'}\,,
  \hspace{1cm}
T_{S5}=\frac{1}{(2\pi\ell_{s})^{5} g_{s}^{2}\ell_{s}}\,,
\end{equation}

\noindent
are the string and S5-brane tensions, respectively.

\subsubsection{Regularity of the corrected solutions}

From the behavior of the functions near the centers
Eqs.~(\ref{eq:nhbehavior}), we see that the corrected solutions describe
$n_{c}$ regular event horizons as long as the charge parameters are strictly
positive, $q^{a}_{+,-, 0}>0$. When the corrections are taken into account,
though, even if we take all these parameters to be positive, there exists the
possibility that the functions $\mathcal{Z}_{+}$ and $\mathcal{Z}_{0}$ vanish
at some points outside the event horizons, giving rise to naked (curvature)
singularities. However, we can argue that in the solutions we are considering in this
paper, all eventual curvature singularities that might appear must be spurious.\footnote{Note that this claim does not apply to the conical singularities (struts) we are also discussing in the paper.} The reason is simply that they are not present in the zeroth-order
solutions. Therefore, if they appear in the corrected solutions, it must be in
a regime of charge parameters where the perturbative expansion is no longer
justified, as the vanishing of the functions can only happen when the
corrections cancel the leading-order contributions, which tells us that they
are equally important. This can be seen very clearly in the single-center
solutions. In this case, the corrections are given by \cite{Cano:2021nzo}

\begin{subequations}
\begin{align}
  \mathcal{Z}_{+}
  & =
    1+\frac{q_{+}}{\rho^{2}}
    +\frac{2\epsilon q_{+}\alpha' \left(\rho^{2}+q_{0}+q_{-}\right)}{q_{0}
    \left(\rho^{2}+q_{-}\right)\left(\rho^{2}+q_{0}\right)}
    +\mathcal{O}(\alpha^{\prime\, 2})\,,
  \\
  & \nonumber \\
  \mathcal{Z}_{0}
  & =
    1+\frac{q_{0}}{\rho^{2}}
    -\alpha'\frac{\rho^{2}+2q_{0}}{\left(\rho^{2}+q_{0}\right)^{2}}
    +\mathcal{O}(\alpha^{\prime\, 2})\,.
\end{align}
\end{subequations}

\begin{figure}[ht!]
\begin{center}
\includegraphics[scale=0.9,clip]{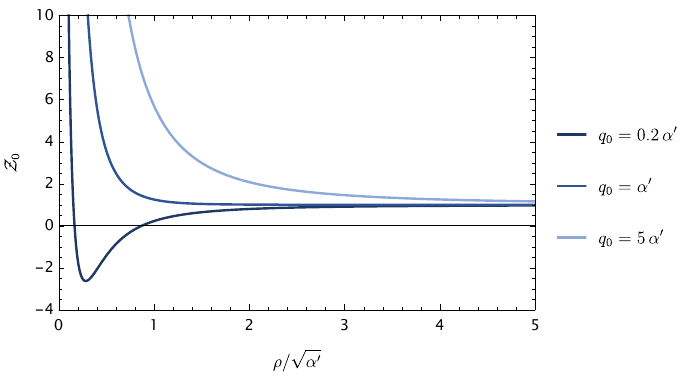}
\caption{Plot of the function $\mathcal{Z}_{0}$ for the solution with a single
  center $n_{c}=1$ and for several values of the charge
  $q_{0}/\alpha'=\{0.2, 1, 5\}$. As we can see, when the value of $q_{0}$ is
  sufficiently small, (spurious) naked singularities appear.}
\label{fig:Z0}
\end{center}
\end{figure}

\begin{figure}[ht!]
\begin{center}
\includegraphics[scale=0.9,clip]{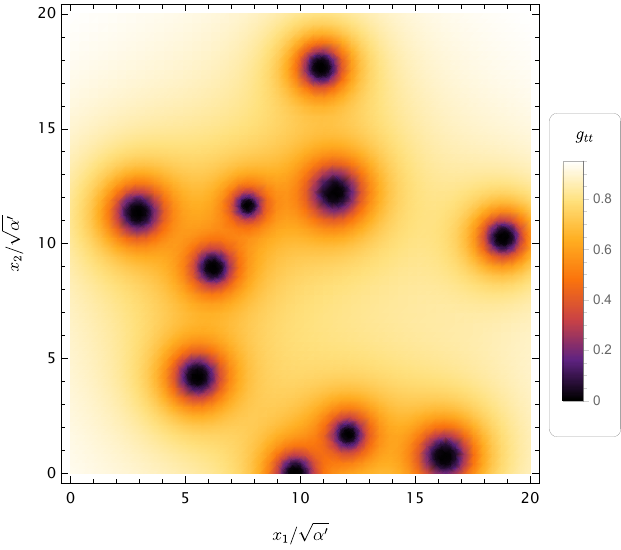}
\caption{Plot of the $g_{tt}$ component of the metric in the Einstein frame
  Eq.~(\ref{eq:esametrica}) for a planar configuration of 3-charged extremal
  BHs in 5 dimensions.  $x_{1}$ and $x_{2}$ are the same of
  Eq.~(\ref{eq:rhoa}). The charges and the positions are positive and randomly
  picked such that the average distance among the BHs is 5 times the average
  value of the charges in units of $\alpha'$.}
\label{fig:P02}
\end{center}
\end{figure}

\begin{figure}[ht!]
\begin{center}
\includegraphics[scale=0.9,clip]{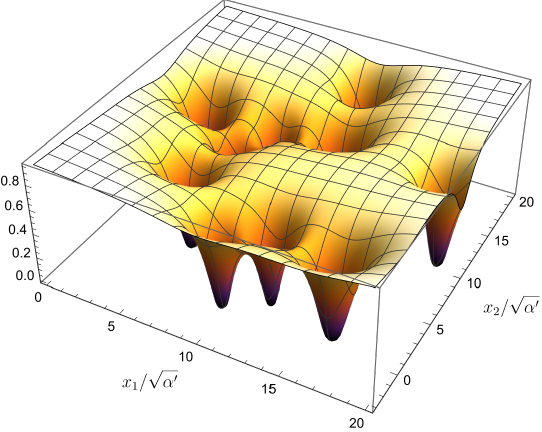}
\caption{3-dimensional version of the plot in Fig.~\ref{fig:P02}. In this
  figure we do not have infinite throats because the coordinates we are using
  are defined outside of the BHs horizons where $g_{tt}=0$.}
\label{fig:P03}
\end{center}
\end{figure}

\noindent
The function $\mathcal{Z}_{+}$ is always strictly positive if
$q_{+,-,0}>0$. However, the function $\mathcal{Z}_{0}$ can vanish and change
sign, which is far worse. However, this only happens if $q_{0}\ll \alpha'$
(see Fig.~\ref{fig:Z0}), in which case the perturbative expansion is no longer
justified. Hence, naked singularities of the corrected solutions only arise
for values of the parameters for which the perturbative expansion breaks down
and we should not worry about them. A graphic depiction of a particular
solution is givenin Figs.~\ref{fig:P02} and \ref{fig:P03}

\begin{figure}[ht!]
	\begin{center}
		\includegraphics[clip]{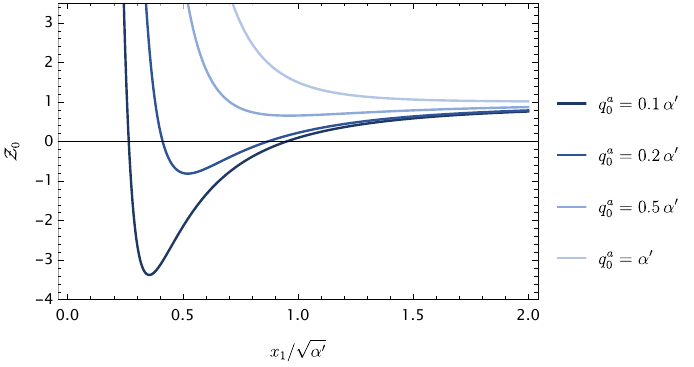}
		\caption{Plot of the function $\mathcal{Z}_{0}$ for the
                  solution with multiple centers $n_{c}=5$ along the plane
                  $x^{3} = x^{4} = 0$. The plane spanned by $x^{1}$ and $x^{2}$ does
                  not contain any of the centers. The centers are placed in
                  the vertexes of a 5-cell.}
		\label{fig:multiZ0}
	\end{center}
\end{figure}
\begin{figure}[ht!]
	\begin{center}
		\includegraphics[scale=0.9,clip]{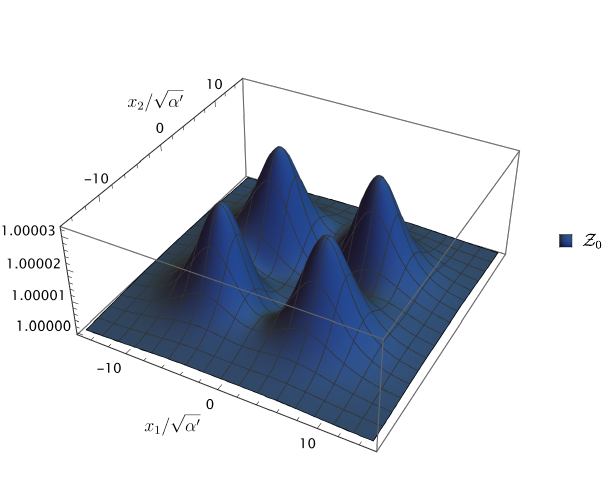}
		\caption{Plot of the function $\mathcal{Z}_{0}$ for the
                  solution with multiple centers $n_{c}=5$ along a direction
                  passing through a center for several values of the charge of
                  such center $q_{0}^{a}/\alpha'=\{0.1, 0.2, 0.5,1\}$. As we can
                  see, when the value of $q_{0}$ is much  smaller than $\alpha'$,
                  (spurious) naked singularities appear. The centers are
                  placed in the vertexes of a 5-cell.}
		\label{fig:singleZ0}
	\end{center}
\end{figure}

In the multi-center case, we have exactly the same situation. In any case, it
is, of course, important to check that the we have found solutions are free of
these pathologies when we take the charges to be large as compared to
$\alpha'$. We have done so numerically for several multi-center configurations
and we have found no singularities whenever the charges are larger than
$\ell_{s}$, in agreement with our general argument: even if we consider a
complicated configuration (see Fig.~\ref{fig:multiZ0}), approaching a center
the corrections to $\mathcal{Z}_{+, \,0}$ reduce to those of a single BH (see
Fig.~\ref{fig:singleZ0}).

\subsubsection{Thermodynamic properties of the corrected black holes}

The ADM mass of the $\alpha'$-corrected five-dimensional black holes can be
straightforwardly computed from the asymptotic expansion of $tt$ component of
the metric. The result can be expressed in several ways,

\begin{equation}
  \label{eq:correctedmass5dbhs}
  \begin{aligned}
    M
    & =
\frac{n}{R_{z}}+\frac{R_{z}}{\alpha'}w
+\frac{R_{z}}{g_{s}^{2}\alpha'}N
+\frac{2\epsilon}{R_{z}}\sum_{a}\frac{n^{a}}{N^{a}}
-\frac{n_{c}R_{z}}{g_{s}^{2}\alpha'}\\
& \\
& =
\frac{|\mathcal{Q}_{+}|}{R_{z}}+\frac{R_{z}|\mathcal{Q}_{-}|}{\alpha'}
+\frac{R_{z}|\mathcal{Q}_{0}|}{g_{s}^{2}\alpha'}\,,
\end{aligned}
\end{equation}

\noindent
which highlight different aspects of the corrections to the mass, as we are
going to discuss. First, observe that we can associate a mass to the
a$^{\rm{th}}$ black hole given by\footnote{These individual masses can be
  found by setting $n_{c}=1$ in Eq.~(\ref{eq:correctedmass5dbhs}).}

\begin{equation}
  M^{a}
  =
  \frac{n^{a}}{R_{z}}\left(1+\frac{2\epsilon}{N^{a}}\right)
  +\frac{R_{z} w^{a}}{\alpha'}
  +\alpha' \left(N^{a}-1\right)\,,
\end{equation}

\noindent
and that it is still true that the total mass is given by the sum of these
putative individual black-hole masses, $M=\sum_{a}M^{a}$. Again, this is
telling us that there is a cancellation of the interaction energies between
the black holes, which is precisely what allows them to be in static
equilibrium. However, contrarily to what happens at zeroth order, the total
mass not only depends on the total amounts of winding, momentum and S5-branes,
but also on the particular distribution of these objects among the different
centers.

In order to discuss the corrections to the mass, we have to specify the
quantities that we keep fixed. There are two natural possibilities: either we
fix the asymptotic (Maxwell) charges, $\mathcal{Q}_{+,-,0}$, or the total number
of stringy sources, $n, w$ and $N$. This distinction is necessary because
these quantities do not coincide when taking into account $\alpha'$
corrections, as we have just seen in Eq.~\eqref{eq:charges5d}. In fact, the
correction to the mass just vanishes when the asymptotic charges are kept
fixed, as reported in \cite{Cano:2021nzo}. In contrast, when fixing $n, w$ and
$N$, one finds that the correction to the mass is given by

\begin{equation}
  \delta M|_{n, w, N}
  =
  \frac{2\epsilon}{R_{z}}\sum_{a}\frac{n^{a}}{N^{a}}
  -\frac{n_{c}R_{z}}{g_{s}^{2}\alpha'}
  =
  \frac{1}{R_{z}}\left[2\epsilon\sum_{a} \frac{n^{a}}{N^{a}}
    -\frac{n_{c} k^{2}_{\infty}}{g_{s}^{2}}\right]\,.
\end{equation}

As we can see, the correction in the non-supersymmetric case ($\epsilon=0$) is
always negative and, furthermore, it decreases linearly with the number of
centers. This means that the fragmentation of these bound states is
energetically favored in the non-supersymmetric case. We will come back to
this issue in Section~\ref{sec-discussion}. In the supersymmetric case, in
turn, the correction does not have a definite sign.

The entropy of these black holes can be obtained by means of the entropy
formula derived in \cite{Elgood:2020nls}\footnote{As explained in that
  reference, in general, the Iyer-Wald formula derived in \cite{Iyer:1994ys}
  using the formalism of \cite{Lee:1990nz,Wald:1993nt} is not valid in
  presence of matter fields because the vast majority of them have gauge
  freedoms that were not correctly taken into account. In particular, in the
  case of the first-order in $\alpha'$ heterotic superstring effective action,
  the entropy formula derived using the Iyer-Wald prescription is not gauge
  invariant (it is frame-dependent). In certain frames, the entropy formula
  derived using the Iyer-Wald prescription and the entropy formula derived in
  \cite{Elgood:2020nls} take the same form, except for a factor of 2 in one of
  the terms. The factor of 2 that occurs in the entropy formula derived in
  \cite{Elgood:2020nls} leads to an entropy that satisfies the first law of
  black-hole thermodynamics in the $\alpha'$-corrected non-extremal
  Reissner-Nordstr\"om black hole \cite{Cano:2019ycn}.} based on the results
of \cite{Elgood:2020svt,Elgood:2020mdx}, as it has recently been done in
\cite{Cano:2021nzo} for single black-hole solutions. Since this entropy is
just an integral over the event horizon, it is just the sum of all the entropies
of the individual black holes, and we can simply use the result in
\cite{Cano:2021nzo} to get

\begin{equation}
  S_{\rm W}=\sum_{a}S^{a}_{\rm W}\,, \hspace{1cm} S^{a}_{\rm W}
  =
  2\pi
\sqrt{n^{a}w^{a}N^{a}}\left(1+\frac{2\epsilon}{N^{a}}\right)\,.
\end{equation}

\noindent
As we can see, the correction to the entropy for fixed $n^{a}, w^{a}$ and
$N^{a}$ vanishes in the non-supersymmetric case, in agreement with the results
of \cite{DominisPrester:2008ynb,Prester:2010cw}. The result in the
supersymmetric case coincides with previous results in the literature, see
\cite{Castro:2008ys,DominisPrester:2008ynb,Prester:2010cw,Faedo:2019xii,Elgood:2020xwu}.
 
\subsection{Multi-Center black-hole solutions in four dimensions}

Plugging Eqs.~(\ref{eq:Zs4d}) in Eqs.~(\ref{eq:Zpcorrected}) and
(\ref{eq:Z0corrected}) and choosing the harmonic functions $\mathcal{H}_{0,+}$
with the same prescription as in the five-dimensional case, we obtain that the
corrections to $\mathcal{Z}_{+}$ and $\mathcal{Z}_{0}$ are the
following,\footnote{The expressions for the harmonic functions in this case
  are
  $\mathcal{H}_{+}=\sum_{a}\frac{\epsilon
    q^{a}_{+}}{2q^{a}_{0}q^{a}_{\mathcal{H}}r_{a}}$ and
  $\mathcal{H}_{0}=-\sum_{a}\frac{1}{2q^{a}_{\mathcal{H}}r_{a}}$.}

\begin{subequations}
  \begin{align}
    \mathcal{Z}_{+}
    & =
      \mathcal{Z}_{+}^{(0)}
      -\frac{\epsilon \alpha'}{2}
      \left[
      \mathcal{Z}_{0}^{(0)\,-1}\mathcal{Z}_{-}^{-1}\mathcal{H}^{-1}
      \sum_{a,b}\frac{
      q^{a}_{+}q^{b}_{-}n_{a}^{m}n_{b}^{m}}{r^{2}_{a}r^{2}_{b}}
      -\sum_{a}\frac{q^{a}_{+}}{q^{a}_{0}q^{a}_{\mathcal{H}}r_{a}}
      \right]
      +\mathcal{O}(\alpha^{\prime\, 2})\,,
    \\
    & \nonumber \\
    \mathcal{Z}_{0}
    & =
      \mathcal{Z}_{0}^{(0)}
      +\frac{\alpha'}{4}\left[
      \mathcal{Z}_{0}^{(0)\,-2}\mathcal{H}^{-1}
      \sum_{a,b}\frac{q^{a}_{0}q^{b}_{0}n_{a}^{m}n_{b}^{m}}{r^{2}_{a}r^{2}_{b}}
      +\mathcal{H}^{-3}\sum_{a,b}\frac{q^{a}_{\mathcal{H}}q^{b}_{\mathcal{H}}n_{a}^{m}n_{b}^{m}}{r^{2}_{a}r^{2}_{b}}
      -\sum_{a}\frac{2}{q^{a}_{\mathcal{H}}r_{a}}
      \right]+\mathcal{O}(\alpha^{\prime\, 2})\,,
  \end{align}
\end{subequations}

\noindent
where, now, $n^{m}_{a}\equiv (x^{m}-x^{m}_{a})/r_{a}$ and $m=1,2,3$.

The behavior of all the functions that determine the four-dimensional
solutions near the black-hole horizons ($r_{a}\rightarrow 0$) is,

\begin{equation}
\mathcal{Z}_{+}\sim\frac{g_{s}^{2}\alpha^{\prime\, 2}n^{a}}{2R^{2}_{z}R_{w}r_{a}}\,,
\hspace{1cm}
\mathcal{Z}_{-}\sim\frac{g_{s}^{2}\alpha' w^{a}}{2R_{w}r_{a}}\,,
\hspace{1cm}
\mathcal{Z}_{0}\sim\frac{\alpha' N^{a}}{2R_{w}r_{a}}\,,
\hspace{1cm}
\mathcal{H}\sim \frac{R_{w}W^{a}}{2r_{a}}\,,
\end{equation}

\noindent
where we have used Eqs.~(\ref{eq:sources4d}) and (\ref{eq:KKquantization}).
Asymptotically, we find the following behavior

\begin{equation}
\mathcal{Z}_{+}\sim 1+\frac{g_{s}^{2}\alpha^{\prime\, 2}|\mathcal{Q}_{+}|}{2R^{2}_{z}R_{w}r}\,,
\hspace{.4cm}
\mathcal{Z}_{-}\sim1+\frac{g_{s}^{2}\alpha' |\mathcal{Q}_{-}|}{2R_{w}r}\,,
\hspace{.4cm}
\mathcal{Z}_{0}\sim1+\frac{\alpha' |\mathcal{Q}_{0}|}{2R_{w}r}\,,
\hspace{.4cm}
\mathcal{H}\sim 1+\frac{R_{w}|\mathcal{Q}_\mathcal{H}|}{2r}\,,
\end{equation}

\noindent
where we have defined

\begin{equation}\label{eq:charges4d}
  \mathcal{Q}_{+}=n+2\epsilon\sum_{a}\frac{n^{a}}{N^{a}W^{a}} \,,
  \hspace{.5cm}
  \mathcal{Q}_{-}=\left(2\epsilon-1\right)w\,,
  \hspace{.5cm}
  \mathcal{Q}_{0}=-\left(N-\sum_{a}\frac{2}{W^{a}}\right)\,,
  \hspace{.5cm} \mathcal{Q}_\mathcal{H}=W\,,
\end{equation}

\noindent
which again coincide with the asymptotic charges of the solution.

\subsubsection{Thermodynamic properties of the corrected solutions}

The total mass of the four-dimensional black-hole solution is given by

\begin{equation}
\begin{aligned}
  M & =
\frac{n}{R_{z}}+\frac{R_{z}w}{\alpha'}
+\frac{R_{z}N}{g_{s}^{2}\alpha'}
+\frac{R^{2}_{w} R_{z}W}{g_{s}^{2}\alpha^{\prime\,2}}
+\frac{2\epsilon}{R_{z}}\sum_{a}\frac{n^{a}}{N^{a}W^{a}}
-\frac{2R_{z}}{g_{s}^{2}\alpha'}\sum_{a}\frac{1}{W^{a}}
\\
& \\
& =
\frac{|\mathcal{Q}_{+}|}{R_{z}}+\frac{R_{z}|\mathcal{Q}_{-}|}{\alpha'}
+\frac{R_{z}|\mathcal{Q}_{0}|}{g_{s}^{2}\alpha'}
+\frac{R^{2}_{w} R_{z} |\mathcal{Q}_\mathcal{H}|}{g_{s}^{2}\alpha^{\prime\,2}}\,.
\end{aligned}
\end{equation}

\noindent
It displays analogous features to the mass of the five-dimensional
ones. First, the total mass is again the sum of the putative individual masses
of each black hole, given by

\begin{equation}
  M^{a}
  =
  \frac{n^{a}}{R_{z}}\left(1+\frac{2\epsilon}{N^{a}W^{a}}\right)+\frac{R_{z}w^{a}}{\alpha'}
+\frac{R_{z}\left(N^{a}-\frac{2}{W^{a}}\right)}{g_{s}^{2}\alpha'}
+\frac{R^{2}_{w} R_{z}W^{a}}{g_{s}^{2}\alpha^{\prime\,2}}\,.
\end{equation}

Again, this fact can be interpreted as directly related to the no-force
condition between the black holes. In addition to this, we observe that the
correction to the mass also vanishes when we keep the asymptotic charges
constant. Instead, when we keep the parameters $n, w, N$ and $W$ constant, we
obtain a non-vanishing correction,

\begin{equation}
  \delta M|_{n, w, N, W}
  =
  \frac{2}{R_{z}}\left[\sum_{a}\frac{\epsilon
n^{a}}{N^{a}W^{a}}-\frac{k^{2}_{\infty}}{g_{s}^{2}}\sum_{a}
\frac{1}{W^{a}}\right] \,,
\end{equation}

\noindent
which is, again, negative in the non-supersymmetric case and decreases with
the number of centers.

The Wald entropy can be computed as in the five-dimensional case, making use
of the results of \cite{Cano:2021nzo}, and it has the value

\begin{equation}
  S_{\rm W}
  =
  2\pi
\sum_{a}\sqrt{n^{a}w^{a}N^{a}W^{a}}\left(1+\frac{2\epsilon}{N^{a}W^{a}}\right)\,.
\end{equation}

\section{Discussion}
\label{sec-discussion}

Our main results can be summarized as follows: we have computed the
first-order $\alpha'$ corrections to multi-center black-hole solutions in five
and four dimensions which provide effective descriptions of bound states of
fundamental strings, momentum waves, solitonic 5-branes and KK monopoles. We
have shown that the first-order $\alpha'$ corrections do not introduce any
singularities and that, therefore, the equilibrium of forces between them is
preserved.

We have also seen how the relation between the asymptotic charges and the
numbers of fundamental objects is altered by the $\alpha'$ corrections, see
Eqs.~(\ref{eq:charges5d}) and (\ref{eq:charges4d}). As a consequence, we have
seen that the mass can have corrections when $n$, $w$, $N$ and $W$ are kept
fixed even if the corrections for the asymptotic charges vanish. Interestingly
enough, we have observed that when the total number of fundamental objects is
kept fixed, the masses of the non-supersymmetric solutions receive negative
corrections that decrease with the number of centers, indicating us that
black-hole fragmentation is energetically favored. The question now is
whether this process is allowed or not.

Clearly, a necessary condition for the fragmentation process to be allowed is
that the conserved charges of the initial and final configurations are
identical. At the two-derivative level, the conserved charges are proportional
to the total numbers of the different fundamental objects, which means that
these numbers must not change. The number of centers $n_{c}$ can, in
principle, change, but it does not appear explicitly in the entropy formula at
this order. Then, at this order, the fragmentation is, in principle, allowed,
but entropically disfavored.

As already mentioned, in presence of $\alpha'$ corrections (which introduce
Chern-Simons terms), one can define several notions of charge and not all of
them are necessarily conserved \cite{Marolf:2000cb}. Therefore, the first
thing we have to do is to figure out which notions of charge are conserved and
which are not.  A thorough analysis of all the possible notions of charge,
their physical interpretation and their properties in this context requires
much more work and will be carried out elsewhere \cite{kn:ORZ}. Thus, here we
will just focus on one of them, the solitonic 5-brane charge. The presence of
S5 branes modifies the Bianchi identity of the Kalb-Ramond 2-form $\hat{B}$ as
follows:

\begin{equation}\label{eq:Bianchi}
  \frac{\hat{g}_{s}^{2}}{16 \pi {G}^{(10)}_{\rm N}}
  \left[d\hat{H}-\frac{\alpha'}{4}\hat{R}_{(-)}{}^{\hat{a}}{}_{\hat{b}}\wedge
    \hat{R}_{(-)}{}^{\hat{b}}{}_{\hat{a}}\right]
  =
  \hat{\star} \hat{J}_{S5}\,.
\end{equation}

The current $J_{S5}$ describes the coupling of external sources (S5 branes) to
the magnetic dual of the KR 2-form. Following \cite{Marolf:2000cb}, we refer
to it as the brane-source current. By definition, it is localized, which in
this context means that it vanishes whenever the sourceless (supergravity)
equations of motion are satisfied. For instance, in the five-dimensional
configurations we have studied, the brane-source current associated to S5
branes is given by 

\begin{equation}
  \hat{\star}\hat{J}_{S5}
  =
  -\hat{g}_{s}^{2}T_{S5}\sum_{a} N^{a} \star_{4} \delta^{(4)}(x-x_{a})\,,
  \hspace{.5cm}
  {\text{where}}
  \hspace{0.5cm}
  \int_{\mathbb {E}^{4}} \star_{4} \,\delta^{(4)}(x-x_{a})=1\,,
\end{equation}

\noindent
since it is precisely at the centers where
Eqs.~\eqref{eq:harmonicityconditions} are not satisfied. Therefore, the
brane-source charge, defined as the integral of $\hat\star \hat{J}_{S5}$, is
proportional to (minus) the total number of S5 branes,

\begin{equation}
  \int_{\mathbb{E}^{4}} \hat{\star}\hat{J}_{S5}
  =
  -\hat{g}_{s} ^{2} T_{S5}\sum_{a}N^{a}
  =
  -\hat{g}_{s} ^{2} T_{S5}N\,.
\end{equation}

As explained in \cite{Marolf:2000cb}, brane-source charges are not conserved
quantities in presence of Chern-Simons terms. The heterotic case is,
however, a bit peculiar because, when taking a exterior derivative in
\eqref{eq:Bianchi}, we arrive to

\begin{equation}
  d\hat\star\hat{J}_{S5}
  =
  \frac{\hat{g}_{s}^{2} \alpha'}{32 \pi G^{(10)}_{\rm N}}
  \hat{\mathcal{D}}_{(-)} \hat{R}_{(-)\, \hat{a} \hat{b}}\wedge \hat{R}_{(-)}{}^{\hat{a}\hat{b}}=0\,,
\end{equation}

\noindent
if the Bianchi identity of the curvature tensor of the torsionful spin
connection $\hat{R}_{(-)}{}^{\hat{a} \hat{b}}$, defined in
Eq.~(\ref{eq:curvaturetorsionfulspinconnection}), is not modified by the
presence of sources, that is

\begin{equation}
  \hat{\cal D}_{(-)}\hat{R}_{(-)}{}_{\hat{a} \hat{b}}
  =
  \hat {\cal D}\hat{R}_{\hat{a}\hat{b}}=0\,.
\end{equation}

This implies that the total number of 5-branes, $N$, must remain constant in
the fragmentation process. The same happens with the Maxwell solitonic 5-brane
charge, which in the five-dimensional case is given by
$\mathcal{Q}_{0}=-\left(N-n_{c}\right)$, see Eq.~\eqref{eq:charges5d}. Hence,
it is evident that the fragmentation is forbidden if both Maxwell and
brane-source charges are conserved. An analogous analysis in the
four-dimensional case yields the same conclusion.

\section*{Acknowledgments}

We are pleased to thank Roberto Emparan, Pedro F. Ram\'irez, Stefano Massai, Luca Martucci and Jose Ju\'an Fern\'andez-Melgarejo for very valuable discussions. 
This work has been supported in part by the MCIU, AEI,
FEDER (UE) grant PGC2018-095205-B-I00 and by the Spanish Research Agency
(Agencia Estatal de Investigaci\'on) through the grant IFT Centro de
Excelencia Severo Ochoa SEV-2016-0597.  AR is supported by a postdoctoral
fellowship associated to the MIUR-PRIN contract 2017CC72MK003.  The work of MZ
was supported by the fellowship LCF/BQ/DI20/11780035 from ``la Caixa''
Foundation (ID 100010434). TO wishes to thank M.M.~Fern\'andez for her
permanent support.

\appendix

\section{The heterotic superstring effective action}
\label{app-action}

The $\alpha'$ corrections that appear in the effective action of the heterotic
superstring were first studied in
\cite{Gross:1986mw,Metsaev:1987zx,Bergshoeff:1989de}, see also \cite{Chemissany:2007he,Baron:2017dvb,Baron:2018lve, Marques:2015vua, Bedoya:2014pma} for more recent studies.  Here we are
following \cite{Bergshoeff:1989de}, adapting their results to the conventions
of \cite{Ortin:2015hya}. Working with a consistent truncation in which all the
gauge fields are trivialized, we have that the effective action of the
heterotic string that includes the first-order $\alpha$'
corrections\footnote{Some terms of second order in $\alpha'$ are implicitly
  included in this expression, in order to give it a more convenient form, but
  they should be consistently ignored.} is given by

\begin{equation}
  \label{eq:action}
  \hat{S}
  =
  \frac{\hat{g}_{s}^{2}}{16\pi G_{\rm N}^{(10)}} \int
  d^{10}x\sqrt{|\hat{g}|}\, e^{-2\hat{\phi}}\,
  \left[
    \hat{R} -4(\partial\hat{\phi})^{2} +\frac{1}{2\cdot 3!}\hat{H}^{2}
    -\frac{\alpha'}{8}\hat{ R}_{(-)\, \hat{\mu}\hat{\nu}}{}^{\hat{a}}{}_{\hat{b}}
\hat{R}_{(-)}{}^{\hat{\mu}\hat{\nu}\, \hat{b}}{}_{\hat{a}} \right]\,,
\end{equation}

\noindent
where

\begin{equation}
  \label{eq:curvaturetorsionfulspinconnection}
  \hat{R}_{(-)}{}^{\hat{a}}{}_{\hat{b}}
  =
  d\hat{\Omega}_{(-)}{}^{\hat{a}}{}_{\hat{b}}
  -\hat{\Omega}_{(-)}{}^{\hat{a}}{}_{\hat{c}} \wedge \hat{\Omega}_{(-)}{}^{\hat{c}}{}_{\hat{b}}\,,
\end{equation} 

\noindent
is the curvature of the torsionful spin connection,

\begin{equation}
  \hat{\Omega}_{(-)}{}^{\hat{a}}{}_{\hat{b}}
  \equiv
  \hat{\omega}^{\hat{a}}{}_{\hat{b}}
  -\frac{1}{2}\hat{H}^{(0)}_{\hat{c}}{}^{\hat{a}}{}_{\hat{b}}\,\hat{e}^{\hat{c}}\,,
\end{equation}

\noindent
and, in their turn, $\hat{\omega}^{\hat{a}}{}_{\hat{b}}$ is  the Levi-Civita spin
connection and

\begin{equation}
  \hat{H}^{(0)}
  =
  d\hat{B}\,,
\end{equation}

\noindent
is the zeroth-order the 3-form field strength of the Kalb-Ramond 2-form
$\hat{B}$. The first-order field strength is given by

\begin{equation}
\label{def:kalb-ramond}
\hat{H}
=
d\hat{B}+ \frac{\alpha'}{4} \hat{\Omega}^{\mathrm{L}}_{(-)}\,,
\end{equation}

\noindent 
where

\begin{equation}
  \label{def:chern-simons}
  \hat{\Omega}^{\mathrm{L}}_{(-)}
  =
  d\hat{\Omega}_{(-)}{}^{\hat{a}}{}_{\hat{b}}\wedge
  \hat{\Omega}_{(-)}{}^{\hat{b}}{}_{\hat{a}}
  -\frac{2}{3} \hat{\Omega}_{(-)}{}^{\hat{a}}{}_{\hat{b}} \wedge
   \hat{\Omega}_{(-)}{}^{\hat{b}}{}_{\hat{c}} \wedge \hat{\Omega}_{(-)}{}^{\hat{c}}{}_{\hat{a}} \,,
\end{equation}

\noindent
is the Chern-Simons 3-form of
$\hat{\Omega}_{(-)}{}^{\hat{a}}{}_{\hat{b}}$. Thus, $\hat{H}$ satisfies the Bianchi
identity

\begin{equation}\label{eq:bianchi} 
  d\hat{H}
  =
  \frac{\alpha'}{4}\hat{R}_{(-)}{}^{\hat{a}}{}_{\hat{b}}\wedge\hat{R}_{(-)}{}^{\hat{b}}{}_{\hat{a}} \,.
\end{equation}

\subsection{Equations of motion}

In order to derive the equations of motion, it is highly convenient to use the
lemma proven in \cite{Bergshoeff:1989de}, which states that the variation of
the action with respect to the torsionful spin connection,
$\delta S/\delta {\hat\Omega}_{(-)}{}^{\hat{a}}{}_{\hat{b}}$, yields terms of
order $\mathcal{O}(\alpha^{\prime\, 2})$ when evaluated on-shell. Since we are
ignoring such higher-order terms, we can simply obtain the equations of motion
by varying the action only with respect to explicit occurrences of the fields
(\textit{i.e.}, those which do not occur through the torsionful spin
connection). Doing so, one obtains

\begin{subequations}
\begin{align}
  \label{eq:eq1}
  \hat{R}_{\hat{\mu}\hat{\nu}} -2\hat{\nabla}_{\hat{\mu}}\partial_{\hat{\nu}}\hat{\phi}
+\frac{1}{4}\hat{H}_{\hat{\mu}\hat \rho\hat \sigma}\hat{H}_{\hat{\nu}}{}^{\hat \rho\hat \sigma}
-\frac{\alpha'}{4}\hat{R}_{(-)\, \hat{\mu}\hat{\rho}}{}^{\hat{a}}{}_{\hat{b}}
  \hat{R}_{(-)\, \hat{\nu}}{}^{\hat{\rho}\,\hat{b}}{}_{\hat{a}}
  & =
    \mathcal{O}(\alpha^{\prime\, 2})\,,
  \\
                     &  \nonumber \\
  \label{eq:eq2}
  (\partial \hat{\phi})^{2} -\frac{1}{2}\hat{\nabla}^{2}\hat{\phi}
  -\frac{1}{4\cdot 3!}\hat{H}^{2}
  +\frac{\alpha'}{32}\hat{R}_{(-)\,\hat{\mu}\hat{\nu}}{}^{\hat{a}}{}_{\hat{b}}
  \hat{R}_{(-)}{}^{\hat{\mu}\hat{\nu}\,\hat{b}}{}_{\hat{a}}
  & =
    \mathcal{O}(\alpha^{\prime\, 2})\,,
  \\
                   &  \nonumber \\
  \label{eq:eq3}
  d\left(e^{-2\hat{\phi}}\star\!\hat{H}\right)
                   & = 
\mathcal{O}(\alpha^{\prime\, 2})\,.
\end{align}
\end{subequations}

\section{Torus compactification}
\label{app:dimensionalreductiontorus}

The reduction of the heterotic superstring effective action over a torus to
first order in $\alpha'$ has been carried out in
Ref.~\cite{Ortin:2020xdm}. Here, however, we only need to know the relation
between the lower- and higher-dimensional fields at leading order in the
$\alpha'$ expansion:\footnote{Here we are using the first letters of the Greek
  alphabet for the internal indices instead of the Latin letters $m,n,\cdots$,
  to avoid possible confusions with the indices of the hyper-K\"ahler
  space. We are also using an unconventional $C$ for some of the fields
  descending from the Kalb-Ramond 2-form.}

\begin{subequations}
  \label{eq:components10dmetric}
\begin{align}
  \hat{g}_{\mu\nu}
  & =
    g_{\mu\nu} -G_{\alpha\beta}A^{\alpha}{}_{\mu}A^{\beta}{}_{\nu}\,,
  \\
  & \nonumber \\
  \hat{g}_{\mu \alpha}
  & =
    -G_{\alpha \beta}A^{\beta}{}_{\mu}\,,
  \\
  & \nonumber \\
  \hat{g}_{\alpha \beta}
  & =
    -G_{\alpha \beta}\,,
      \\
    & \nonumber \\
      \hat{B}_{\mu \nu}
    & =
      B_{\mu\nu}-A^{\alpha}{}_{[\mu|} C_{\beta\, |\nu]}
      \\
    & \nonumber \\
      \hat{B}_{\mu \alpha}
  & =
    C_{\alpha\, \mu}-B_{\alpha\beta}A^{\beta}{}_{\mu}\,,
  \\
  & \nonumber \\
      \hat{B}_{\alpha \beta}
    & =
      C_{\alpha \beta}\,,
  \\
  & \nonumber \\
  \hat{\phi}
  & =
    \phi +\tfrac{1}{2}\log{\text{det}(G_{\alpha \beta})}\,.
\end{align}
\end{subequations}

The inverse relations are 

\begin{subequations}
  \begin{align}
   g_{\mu\nu}
  & =
   \hat{g}_{\mu\nu}-\hat{g}^{\alpha \beta}\hat{g}_{\mu \alpha }\hat{g}_{\nu \beta}\,,
  \\
  & \nonumber \\
    A^{\alpha }{}_{\mu}
    & =
      \hat{g}^{\alpha \beta}\hat{g}_{\mu \beta}\,,
  \\
  & \nonumber \\
  G_{\alpha \beta}
  & =
    -\hat{g}_{\alpha \beta}\,,
      \\
    & \nonumber \\
    B_{\mu\nu}
    & \equiv
      \hat{B}_{\mu \nu} +\hat{g}^{\alpha \beta}\hat{g}_{\alpha \, [\mu} \hat{B}_{\nu]\,  \beta}\,,
      \\
    & \nonumber \\
    C_{\alpha\, \mu}
    & \equiv
      \hat{B}_{\mu \alpha} +\hat{B}_{\alpha\beta}\hat{g}^{\beta\gamma}\hat{g}_{\mu \gamma}\,,
  \\
  & \nonumber \\
      C_{\alpha\beta}
    & =
      \hat{B}_{\alpha\beta}\,,
  \\
  & \nonumber \\
    \phi
    & =
      \hat{\phi} -\tfrac{1}{2}\log{|\text{det}(\hat{g}_{\alpha\beta})|}\,. 
  \end{align}
\end{subequations}

\section{Curvature components}
\label{app:curvatures}

For convenience, we change coordinates introducing the coordinate $u = t - k_\infty z$. The computations will be performed in the following vielbein basis

\begin{equation} 
  \hat{e}^{+}=\frac{du}{\mathcal{Z}_{-}}\,,
  \hspace{.5cm}
  \hat{e}^{-}=dt-\frac{\mathcal{Z}_{+}}{2}\,du\,,
  \hspace{.5cm}
  \hat{e}^{m}=\mathcal{Z}_{0}^{1/2}\,v^{m}\,,
\end{equation} 

\noindent
where $v^{m}$ is a vierbein of the hyper-K\"ahler metric, namely
$d\sigma^{2}=v^{m}v^{n}\delta_{mn}$. In the following we will use $\partial_m$ to indicate the partial derivative with respect to the flat indexes of the hyper-K\"ahler vielbeins $v^m$, i.e. we use
\begin{equation}
	\partial_m = v_m{}^{\underline{m}}\partial_{\underline{m}}\,.
\end{equation}

\subsection{Torsionful spin connection}

The components of the torsionful spin connection
$\hat{\Omega}_{(-)}{}_{\hat{a} \hat{b}}=\hat{\omega}_{\hat{a}\hat{b}}
-\frac{1}{2}\hat{H}^{(0)}_{\hat{c} \hat{a} \hat{b}}\,\hat{e}^{\hat{c}}$ are given
by

\begin{subequations}
\begin{align}
  \hat{\Omega}_{(-)\, +-}
  & =
    \frac{\epsilon}{\mathcal{Z}_{0}^{1/2}}\partial_{m}\log{\mathcal{Z}_{-}}
    \hat{e}^{m}\,,
  \\
  & \nonumber \\
  \hat{\Omega}_{(-)\, -m}
  & =
    \frac{\epsilon}{\mathcal{Z}_{0}^{1/2}}\partial_{m}\log{\mathcal{Z}_{-}}
    \hat{e}^{+}\,,
  \\
    & \nonumber \\
  \hat{\Omega}_{(-)\, +m}
  & =
    \frac{\mathcal{Z}_{-}}{2\mathcal{Z}_{0}^{1/2}}\partial_{m}\mathcal{Z}_{+} \hat{e}^{+}
    +\frac{1-\epsilon}{\mathcal{Z}_{0}^{1/2}}\partial_{m}\log{\mathcal{Z}_{-}}
    \hat{e}^{-}\,,
  \\
    & \nonumber \\
  \hat{\Omega}_{(-)\, mn}
  & =
    \left[{\tilde{\omega}}_{pmn}+\mathbb{M}^{-}_{mnpq}\partial_{q}\log{\mathcal{Z}_{0}}\right]
    \frac{\hat {e}^{p}}{\mathcal{Z}_{0}^{1/2}}\,,
\end{align}
\end{subequations} 

\noindent
where $\tilde{\omega}_{mn}$ is the hyper-K\"ahler spin connection\footnote{It
  is defined through $dv^{m}+\tilde{\omega}^{m}{}_{n}\wedge v^{n}=0$.} and
$\mathbb{M}^{-}_{mnpq}=\delta_{m[p} \delta_{q]n}-\frac{1}{2}\epsilon_{mnpq}$
are the anti-self-dual $\mathfrak{so}(3)$ subalgebra of of $\mathfrak{so}(4)$.

\subsection{Curvature 2-form}

The non-vanishing components of the curvature 2-form are

\begin{subequations}
  \label{eq:curvature2form}
\begin{align} 
  \hat{R}_{(-)\,-m} & =
                      \frac{\hat{e}^{n}\wedge\hat{e}^{+}}{\mathcal{Z}_{0}}
                      \epsilon\left[\nabla_{n} \partial_{m} \log \mathcal{Z}_{-}
                      -\frac{1}{2}\partial_{m} \log \mathcal{Z}_{-}\partial_{n}\log{\mathcal{Z}_{0}}
                      -{\mathbb
                      M}^{-}_{pmnq}\partial_{p}\log{\mathcal{Z}_{-}}\partial_{q}\log{\mathcal{Z}_{0}}\right]\,,
  \\
  & \nonumber \\
  \hat{R}_{(-)}{}_{+m}
                    & =
                      \frac{\mathcal{Z}_{-}\hat{e}^{n}\wedge
                      \hat{e}^{+}}{2\mathcal{Z}_{0}}
                      \left[\nabla_{n} \partial_{m}
\mathcal{Z}_{+}-\frac{1}{2}\partial_{n}\log \mathcal{Z}_{0} \partial_{m}
\mathcal{Z}_{+}+(\epsilon-1)\partial_{m}\log \mathcal{Z}_{-} \partial_{n}
                        \mathcal{Z}_{0}\right.
                        \nonumber \\
& \nonumber \\
                    &\hspace{.5cm}
                      \left.
                      -\epsilon\partial_{m}
                      \mathcal{Z}_{+}\partial_{n}\log{\mathcal{Z}_{-}}\right]
                      +(1-\epsilon)\frac{e^{n}\wedge e^{-}}{\mathcal{Z}_{0}}
                      \left[\nabla_{n}\partial_{m} \log{\mathcal{Z}_{-}}\right.
                      \nonumber \\
  & \nonumber \\
                    & \hspace{.5cm}
                      \left.
                      -\frac{1}{2}\partial_{m}\log{\mathcal{Z}_{-}}
                      \partial_{n} \log{\mathcal{Z}_{0}}
                      -\mathbb{M}^{-}_{pmnq}\partial_{p}\log{\mathcal{Z}_{-}}\partial_{q}\log{\mathcal{Z}_{0}}\right]\,,\\
                 &  \nonumber \\ 
  \hat{R}_{(-)}{}_{mn}
                    & =
                      \tilde{R}_{mn}+\tilde{F}_{mn}\,,
\end{align}
\end{subequations} 

\noindent
where

\begin{subequations}
\begin{align} 
  \tilde{R}_{mn}
  & =
    d\tilde{\omega}_{mn}+\tilde{\omega}_{mp}\wedge \tilde{\omega}{}_{pn}\,,
  \\
  & \nonumber \\
  \tilde{F}_{mn}
  &=
    d\tilde{A}_{mn}+\tilde{A}_{mp}\wedge \tilde{A}{}_{pn}\,,
\end{align}
\end{subequations} 

\noindent
and

\begin{equation}
  \tilde{A}_{mn}=\mathbb{M}^{-}_{mnpq}\partial_{q}\log{\mathcal{Z}_{0}} \,v^{p}\,.
\end{equation} 

When computing the $mn$ components of the curvature 2-form,
Eq.~(\ref{eq:curvature2form}) it is crucial that the spin connection
$\tilde{\omega}_{mn}$ and the connection $\tilde{A}_{mn}$ satisfy opposite
self-duality relations,

\begin{equation}
  \tilde{\omega}_{mn}
  =
  +\frac{1}{2}\epsilon_{mnpq}\,\tilde{\omega}_{pq}\,,
  \hspace{1cm}
  \tilde{A}_{mn}
  =
  -\frac{1}{2}\epsilon_{mnpq}\tilde{A}_{pq}\,,
\end{equation} 

\noindent
\textit{i.e.}, each of these connections belongs to one of the two orthogonal
subspaces, $\mathfrak{so}_{\pm}(3)$, in which
$\mathfrak{so}(4)=\mathfrak{so}_{+}(3)\oplus\mathfrak{so}_{-}(3)$ splits.

\section{The general $\alpha'$-corrected solution}
\label{app:thesolution}

The $\alpha'$ corrections to the equations of motion and Bianchi identity are
encoded in the so-called T-tensors \cite{Bergshoeff:1989de}, which are given
by

\begin{eqnarray}
  \hat{T}^{(2)}_{\hat{a} \hat{b}}
  &=&
      -\frac{1}{4}\hat{R}_{(-)\,\hat{a}\hat{c}\hat{d}\hat{e}}\hat{R}_{(-)\,\hat{b}}{}^{\hat{c}\hat{d}\hat{e}}\,,
  \\
  & & \nonumber \\
  \hat{T}^{(4)}
  &=&
      \hat{R}_{(-)\,\hat{a}\hat{b}}\wedge \hat{R}_{(-)}{}^{\hat{b}\hat{a}}\,,
  \\
    & & \nonumber \\
  \hat{T}^{(0)}
  &=&
      \hat{T}^{(2)}_{\hat{a}\hat{b}}\eta^{\hat{a}\hat{b}}\,.
\end{eqnarray} 

Using the calculations displayed in Appendix~\ref{app:curvatures}, one can
check that all the T-tensors are equal to those of the supersymmetric case
\cite{Chimento:2018kop} except for $\hat{T}^{(2)}_{++}$, which simply vanishes
in the non-supersymmetric case. Since the left-hand side of the $++$ component
of the Einstein equations is proportional to the Laplacian of
$\mathcal{Z}_{+}$, this function does not receive $\alpha'$ corrections when
$\epsilon=0$. Hence, we find \cite{Chimento:2018kop}

\begin{equation}
  \mathcal{Z}_{+}
  =
  \mathcal{Z}^{(0)}_{+}-\frac{\epsilon\alpha'}{2}
  \frac{\partial_{m}\mathcal{Z}^{(0)}_{+}\partial_{m}\mathcal{Z}^{(0)}_{-}}{\mathcal{Z}_{0}^{(0)}\mathcal{Z}_{-}^{(0)}} +\mathcal{O}(\alpha^{\prime\, 2})\,,
\end{equation}

\noindent
where $\mathcal{Z}^{(0)}_{+,-, 0}$ are the functions that determine the
zeroth-order solution through Eqs.~\eqref{eq:10dsolution}. As discussed in the
main text, the solution for $\mathcal{Z}_{+}$ is only specified up to a
harmonic function, which corresponds to the freedom we have to fix the
boundary conditions.  The correction to the function $\mathcal{Z}_{0}$ can be
found by solving the Bianchi identity (\ref{eq:bianchi}). Given that
$\hat{T}^{(4)}$ does not depend on $\epsilon$, the solution is the same as the
one found in \cite{Chimento:2018kop}, namely

\begin{equation}
  \mathcal{Z}_{0}
  =
  \mathcal{Z}_{0}^{(0)}
  +\frac{\alpha'}{4}\left[\frac{\partial_{m}
      \mathcal{Z}_{0}^{(0)}\partial_{m}\mathcal{Z}_{0}^{(0)}}{(\mathcal{Z}_{0}^{(0)})^{2}}
    +\frac{\partial_{m}
      \mathcal{H}^{(0)}\partial_{m}\mathcal{H}^{(0)}}{(\mathcal{H}^{(0)})^{2}}\right]
  +\mathcal{O}(\alpha^{\prime\, 2})\,.
\end{equation}



\begin{thebibliography}{99}

\bibitem{Ortin:2015hya}
T.~Ort\'{\i}n,
``Gravity and Strings,''
\doi{10.1017/CBO9781139019750}

\bibitem{Strominger:1996sh}
A.~Strominger and C.~Vafa,
``Microscopic origin of the Bekenstein-Hawking entropy,''
Phys. Lett. B \textbf{379} (1996), 99-104
\doi{10.1016/0370-2693(96)00345-0}
[\hepth{9601029} [hep-th]].

\bibitem{Bergshoeff:1989de}
E.~A.~Bergshoeff and M.~de Roo,
``The Quartic Effective Action of the Heterotic String and Supersymmetry,''
Nucl. Phys. B \textbf{328} (1989), 439-468
\doi{10.1016/0550-3213(89)90336-2}

\bibitem{Gross:1986mw}
D.~J.~Gross and J.~H.~Sloan,
``The Quartic Effective Action for the Heterotic String,''
Nucl. Phys. B \textbf{291} (1987), 41-89
\doi{10.1016/0550-3213(87)90465-2}

\bibitem{Metsaev:1987zx}
R.~R.~Metsaev and A.~A.~Tseytlin,
``Order alpha-prime (Two Loop) Equivalence of the String Equations of Motion
and the Sigma Model Weyl Invariance Conditions: Dependence on the Dilaton
and the Antisymmetric Tensor,''
Nucl. Phys. B \textbf{293} (1987), 385-419
\doi{10.1016/0550-3213(87)90077-0}

\bibitem{Chemissany:2007he}
W.~A.~Chemissany, M.~de Roo and S.~Panda,
``alpha'-Corrections to Heterotic Superstring Effective Action Revisited,''
JHEP \textbf{08} (2007), 037
\doi{10.1088/1126-6708/2007/08/037}
[\arxiv{0706.3636} [hep-th]].

\bibitem{Maldacena:1996gb}
J.~M.~Maldacena and A.~Strominger,
``Statistical entropy of four-dimensional extremal black holes,''
Phys. Rev. Lett. \textbf{77} (1996), 428-429
\doi{10.1103/PhysRevLett.77.428}
[\hepth{9603060} [hep-th]].

\bibitem{Johnson:1996ga}
C.~V.~Johnson, R.~R.~Khuri and R.~C.~Myers,
``Entropy of 4-D extremal black holes,''
Phys. Lett. B \textbf{378} (1996), 78-86
\doi{10.1016/0370-2693(96)00383-8}
[\hepth{9603061} [hep-th]].

\bibitem{Cano:2021nzo}
P.~A.~Cano, T.~Ort\'{\i}n, A.~Ruip\'erez and M.~Zatti,
``Non-supersymmetric black holes with \ensuremath{\alpha}' corrections,''
JHEP \textbf{03} (2022), 103
\doi{10.1007/JHEP03(2022)103}
[\arxiv{2111.15579} [hep-th]].

\bibitem{Cano:2018qev}
P.~A.~Cano, P.~Meessen, T.~Ort\'\i{}n and P.~F.~Ram\'\i{}rez,
``$\alpha'$-corrected black holes in String Theory,''
JHEP \textbf{05} (2018), 110
\doi{10.1007/JHEP05(2018)110}
[\arxiv{1803.01919} [hep-th]].

\bibitem{Chimento:2018kop}
  S.~Chimento, P.~Meessen, T.~Ort\'{\i}n,
  P.~F.~Ram\'{\i}rez and A.~Ruip\'erez,
``On a family of $\alpha'$-corrected solutions of the Heterotic Superstring effective action,''
JHEP \textbf{07} (2018), 080
\doi{10.1007/JHEP07(2018)080}
[\arxiv{1803.04463} [hep-th]].

\bibitem{Cano:2018brq}
  P.~A.~Cano, S.~Chimento, P.~Meessen,
  T.~Ort\'{\i}n, P.~F.~Ram\'{\i}rez and A.~Ruip\'erez,
``Beyond the near-horizon limit: Stringy corrections to Heterotic Black Holes,''
JHEP \textbf{02} (2019), 192
\doi{10.1007/JHEP02(2019)192}
[\arxiv{1808.03651} [hep-th]].

\bibitem{Cano:2019oma}
P.~A.~Cano, T.~Ort\'\i{}n and P.~F.~Ram\'{\i}rez,
``On the extremality bound of stringy black holes,''
JHEP \textbf{02} (2020), 175
\doi{10.1007/JHEP02(2020)175}
[\arxiv{1909.08530} [hep-th]].

\bibitem{Cano:2019ycn}
  P.~A.~Cano, S.~Chimento, R.~Linares,
  T.~Ort\'\i{}n and P.~F.~Ram\'\i{}rez,
``$\alpha'$ corrections of Reissner-Nordstr\"om black holes,''
JHEP \textbf{02} (2020), 031
\doi{10.1007/JHEP02(2020)031}
[\arxiv{1910.14324} [hep-th]].

\bibitem{Cano:2022tmn}
P.~A.~Cano, T.~Ort\'\i{}n, A.~Ruip\'erez and M.~Zatti,
``Non-extremal, \ensuremath{\alpha}'-corrected black holes
in 5-dimensional heterotic superstring theory,''
JHEP \textbf{12} (2022), 150
\doi{10.1007/JHEP12(2022)150}
[\arxiv{2210.01861} [hep-th]].

\bibitem{Zatti:2023oiq}
M.~Zatti,
``$\alpha'$ corrections to 4-dimensional non-extremal stringy black holes,''
[\arxiv{2308.12879} [hep-th]].

\bibitem{Bergshoeff:1995cg}
E.~Bergshoeff, B.~Janssen and T.~Ort\'{\i}n,
``Solution generating transformations and the string effective action,''
Class.\ Quant.\ Grav.\  {\bf 13} (1996) 321.
\doi{10.1088/0264-9381/13/3/002}
[\hepth{9506156}].

\bibitem{Elgood:2020xwu}
Z.~Elgood and T.~Ort\'{\i}n,
``T duality and Wald entropy formula in the Heterotic Superstring
effective action at first-order in \ensuremath{\alpha}',''
JHEP \textbf{10} (2020), 097
[erratum: JHEP \textbf{06} (2021), 105]
\doi{10.1007/JHEP10(2020)097}
[\arxiv{2005.11272} [hep-th]].

\bibitem{Hawking:1971vc}
S.~Hawking,
``Black holes in general relativity,''
Commun. Math. Phys. \textbf{25} (1972), 152-166
\doi{10.1007/BF01877517}

\bibitem{Bardeen:1973gs}
J.~M.~Bardeen, B.~Carter and S.~Hawking,
``The Four laws of black hole mechanics,''
Commun.\ Math.\ Phys.\  \textbf{31} (1973), 161-170
\doi{10.1007/BF01645742}

\bibitem{Hawking:1974sw}
S.~Hawking,
``Particle Creation by Black Holes,''
Commun. Math. Phys. \textbf{43} (1975), 199-220
\doi{10.1007/BF02345020}

\bibitem{Racz:1995nh}
I.~Racz and R.~M.~Wald,
``Global extensions of space-times describing
asymptotic final states of black holes,''
Class. Quant. Grav. \textbf{13} (1996), 539-553
\doi{10.1088/0264-9381/13/3/017}
[\grqc{9507055} [gr-qc]].

\bibitem{Christodoulou:1970wf}
D.~Christodoulou,
``Reversible and irreversible transforations in black hole physics,''
Phys. Rev. Lett. \textbf{25} (1970), 1596-1597
\doi{10.1103/PhysRevLett.25.1596}

\bibitem{Christodoulou:1972kt}
D.~Christodoulou and R.~Ruffini,
``Reversible transformations of a charged black hole,''
Phys. Rev. D \textbf{4} (1971), 3552-3555
\doi{10.1103/PhysRevD.4.3552}

\bibitem{Hawking:1971tu}
S.~Hawking,
``Gravitational radiation from colliding black holes,''
Phys. Rev. Lett. \textbf{26} (1971), 1344-1346
\doi{10.1103/PhysRevLett.26.1344}

\bibitem{Wald:1993nt}
R.~M.~Wald,
``Black hole entropy is the Noether charge,''
Phys. Rev. D \textbf{48} (1993) no.8, R3427-R3431
\doi{10.1103/PhysRevD.48.R3427}
[\grqc{9307038} [gr-qc]].

\bibitem{Lee:1990nz}
J.~Lee and R.~M.~Wald,
``Local symmetries and constraints,''
J. Math. Phys. \textbf{31} (1990), 725-743
\doi{10.1063/1.528801}

\bibitem{Wall:2015raa}
A.~C.~Wall,
``A Second Law for Higher Curvature Gravity,''
Int. J. Mod. Phys. D \textbf{24} (2015) no.12, 1544014
\doi{10.1142/S0218271815440149}
[\arxiv{1504.08040} [gr-qc]].

\bibitem{Hollands:2022fkn}
S.~Hollands, \'A.~D.~Kov\'acs and H.~S.~Reall,
``The second law of black hole mechanics in effective field theory,''
JHEP \textbf{08} (2022), 258
\doi{10.1007/JHEP08(2022)258}
[\arxiv{2205.15341} [hep-th]].

\bibitem{Davies:2022xdq}
I.~Davies and H.~S.~Reall,
``Dynamical Black Hole Entropy in Effective Field Theory,''
[\arxiv{2212.09777} [hep-th]].

\bibitem{Iyer:1994ys}
V.~Iyer and R.~M.~Wald,
``Some properties of Noether charge and a proposal for dynamical black hole entropy,''
Phys. Rev. D \textbf{50} (1994), 846-864
\doi{10.1103/PhysRevD.50.846}
[\grqc{9403028} [gr-qc]].

\bibitem{Sen:2005iz}
A.~Sen,
``Entropy function for heterotic black holes,''
JHEP \textbf{03} (2006), 008
\doi{10.1088/1126-6708/2006/03/008}
[\hepth{0508042} [hep-th]].

\bibitem{DominisPrester:2008ynb}
P.~Dominis Prester and T.~Terzic,
``$\alpha'$-exact entropies for BPS and non-BPS extremal
dyonic black holes in heterotic string theory from ten-dimensional supersymmetry,''
JHEP \textbf{12} (2008), 088
\doi{10.1088/1126-6708/2008/12/088}
[\arxiv{0809.4954} [hep-th]].

\bibitem{Faedo:2019xii}
F.~Faedo and P.~F.~Ram\'{\i}rez,
``Exact charges from heterotic black holes,''
JHEP \textbf{10} (2019), 033
\doi{10.1007/JHEP10(2019)033}
[\arxiv{1906.12287} [hep-th]].

\bibitem{Elgood:2020nls}
Z.~Elgood, T.~Ort\'{\i}n and D.~Pere\~n\'{\i}guez,
``The first law and Wald entropy formula of heterotic
stringy black holes at first order in $\alpha'$,''
JHEP \textbf{05} (2021), 110
\doi{10.1007/JHEP05(2021)110}
[\arxiv{2012.14892} [hep-th]].

\bibitem{Tachikawa:2006sz}
Y.~Tachikawa,
``Black hole entropy in the presence of Chern-Simons terms,''
Class. Quant. Grav. \textbf{24} (2007), 737-744
\doi{10.1088/0264-9381/24/3/014}
[\hepth{0611141} [hep-th]].

\bibitem{Elgood:2020svt}
Z.~Elgood, P.~Meessen and T.~Ort\'{\i}n,
``The first law of black hole mechanics in the
Einstein-Maxwell theory revisited,''
JHEP \textbf{09} (2020), 026
\doi{10.1007/JHEP09(2020)026}
[\arxiv{2006.02792} [hep-th]].

\bibitem{Prabhu:2015vua}
K.~Prabhu,
``The First Law of Black Hole Mechanics for Fields
with Internal Gauge Freedom,''
Class.\ Quant.\ Grav.\  {\bf 34} (2017) no.3,  035011.
\doi{10.1088/1361-6382/aa536b}
[\arxiv{1511.00388} [gr-qc]].

\bibitem{Hajian:2015xlp}
K.~Hajian and M.~M.~Sheikh-Jabbari,
``Solution Phase Space and Conserved Charges:
A General Formulation for Charges Associated with Exact Symmetries,''
Phys. Rev. D \textbf{93} (2016) no.4, 044074
\doi{10.1103/PhysRevD.93.044074}
[\arxiv{1512.05584} [hep-th]].

\bibitem{Elgood:2020mdx}
Z.~Elgood, D.~Mitsios, T.~Ort\'\i{}n and D.~Pere\~n\'\i{}guez,
``The first law of heterotic stringy black hole mechanics at
zeroth order in \ensuremath{\alpha}',''
JHEP \textbf{07} (2021), 007
\doi{10.1007/JHEP07(2021)007}
[\arxiv{2012.13323} [hep-th]].

\bibitem{Ortin:2022uxa}
T.~Ort\'{\i}n and D.~Pere\~niguez,
``Magnetic charges and Wald entropy,''
JHEP \textbf{11} (2022), 081
\doi{10.1007/JHEP11(2022)081}
[\arxiv{2207.12008} [hep-th]].

\bibitem{Gibbons:1996af}
G.~W.~Gibbons, R.~Kallosh and B.~Kol,
``Moduli, scalar charges, and the first law of black hole thermodynamics,''
Phys. Rev. Lett. \textbf{77} (1996), 4992-4995
\doi{10.1103/PhysRevLett.77.4992}
[\hepth{9607108} [hep-th]].

\bibitem{Ballesteros:2023iqb}
R.~Ballesteros, C.~G\'omez-Fayr\'en, T.~Ort\'{\i}n and M.~Zatti,
``On scalar charges and black-hole thermodynamics,''
[\arxiv{2302.11630} [hep-th]].

\bibitem{kn:G-FOZ}
  C.~G\'omez-Fayr'en, T.~Ort\'{\i}n and M.~Zatti,
  ``Wald entropy in Kaluza-Klein black holes'',
work  to be submitted.

\bibitem{Marolf:2000cb}
D.~Marolf,
``Chern-Simons terms and the three notions of charge,''
[\hepth{0006117} [hep-th]].

\bibitem{kn:ORZ}
T.~Ort\'{\i}n, A.~Ruip\'erez and M.~Zatti,
  work in progress.

\bibitem{Israel-Kahn}
W. Israel and K. A. Kahn, 
``Collinear particles and Bondi Dipoles in General Relativity,"
Nuovo Cim. \textbf{33}(1964)331.

\bibitem{Costa:2000kf}
M.~S.~Costa and M.~J.~Perry,
``Interacting black holes,''
Nucl. Phys. B \textbf{591} (2000), 469-487
\doi{10.1016/S0550-3213(00)00577-0}
[\hepth{0008106} [hep-th]].

\bibitem{Brill:1963yv}
D.~R.~Brill and R.~W.~Lindquist,
``Interaction energy in geometrostatics,''
Phys. Rev. \textbf{131} (1963), 471-476
\doi{10.1103/PhysRev.131.471}

\bibitem{Hartle:1972ya}
J.~B.~Hartle and S.~W.~Hawking,
``Solutions of the Einstein-Maxwell equations with many black holes,''
Commun. Math. Phys. \textbf{26} (1972), 87-101
\doi{10.1007/BF01645696}

\bibitem{Meessen:2017rwm}
P.~Meessen, T.~Ort\'\i{}n and P.~F.~Ram\'\i{}rez,
``Dyonic black holes at arbitrary locations,''
JHEP \textbf{10} (2017), 066
\doi{10.1007/JHEP10(2017)066}
[\arxiv{1707.03846} [hep-th]].

\bibitem{Lucietti:2020ryg}
J.~Lucietti,
``All Higher-Dimensional Majumdar\textendash{}Papapetrou Black Holes,''
Annales Henri Poincare \textbf{22} (2021) no.7, 2437-2450
\doi{10.1007/s00023-021-01037-0}
[\arxiv{2009.05358} [gr-qc]].

\bibitem{Chrusciel:2006pc}
P.~T.~Chrusciel and P.~Tod,
``The Classification of static electro-vacuum space-times containing an
asymptotically flat spacelike hypersurface with compact interior,''
Commun. Math. Phys. \textbf{271} (2007), 577-589
\doi{10.1007/s00220-007-0191-9}
[\grqc{0512043} [gr-qc]].

\bibitem{Gauntlett:2002nw}
J.~P.~Gauntlett, J.~B.~Gutowski, C.~M.~Hull, S.~Pakis and H.~S.~Reall,
``All supersymmetric solutions of minimal supergravity in five- dimensions,''
Class. Quant. Grav. \textbf{20} (2003), 4587-4634
\doi{10.1088/0264-9381/20/21/005}
[\arxiv{0209114} [hep-th]].

\bibitem{Ortin:1996bz}
T.~Ortin,
``Extremality versus supersymmetry in stringy black holes,''
Phys. Lett. B \textbf{422} (1998), 93-100
\doi{10.1016/S0370-2693(98)00040-9}
[\hepth{9612142} [hep-th]].

\bibitem{Gibbons:1982ih}
G.~W.~Gibbons,
``Antigravitating Black Hole Solitons with Scalar Hair in N=4 Supergravity,''
Nucl. Phys. B \textbf{207} (1982), 337-349
\doi{10.1016/0550-3213(82)90170-5}

\bibitem{kn:Ma} S.\,D.~Majumdar,
        {\sl A Class of Exact Solutions of Einstein's 
        Field Equations},
        {\it Phys.~Rev.}~\textbf{72} (1947) 390--398.

\bibitem{kn:Pa} A.~Papapetrou,
        {\sl A Static Solution of the Equations of the Gravitational 
        Field for an Arbitrary Charge-Distribution},
        {\it Proc.~Roy.~Irish~Acad.}~\textbf{A51} (1947) 191.

\bibitem{VandenBleeken:2008tsa}
D.~Van den Bleeken,
``Multicentered black holes in string theory,''
Ph.D,~Thesis, K.U. Leuven (2008)
        
\bibitem{Maldacena:1996ky}
J.~M.~Maldacena,
``Black holes in string theory,''
[\hepth{9607235} [hep-th]].

\bibitem{Kunduri:2013gce}
H.~K.~Kunduri and J.~Lucietti,
``Classification of near-horizon geometries of extremal black holes,''
Living Rev. Rel. \textbf{16} (2013), 8
\doi{10.12942/lrr-2013-8}
[\arxiv{1306.2517} [hep-th]].

\bibitem{Gibbons:1979zt}
G.~W.~Gibbons and S.~W.~Hawking,
``Gravitational Multi - Instantons,''
Phys. Lett. B \textbf{78} (1978), 430
\doi{10.1016/0370-2693(78)90478-1}

\bibitem{Gibbons:1987sp}
G.~W.~Gibbons and P.~J.~Ruback,
``The Hidden Symmetries of multi-center Metrics,''
Commun. Math. Phys. \textbf{115} (1988), 267
\doi{10.1007/BF01466773}

\bibitem{Prester:2010cw}
P.~Dominis Prester,
``$\alpha'$-Corrections and Heterotic Black Holes,''
[\arxiv{1001.1452} [hep-th]].

\bibitem{Castro:2008ys}
A.~Castro and S.~Murthy,
``Corrections to the statistical entropy of five dimensional
black holes,''
JHEP \textbf{06} (2009), 024
\doi{10.1088/1126-6708/2009/06/024}
[\arxiv{0807.0237} [hep-th]].

\bibitem{Baron:2017dvb}
W.~H.~Baron, J.~J.~Fernandez-Melgarejo, D.~Marqu\'es and C.~N\'u\~nez,
``The Odd story of \ensuremath{\alpha}'-corrections,''
JHEP \textbf{04} (2017), 078
\doi{10.1007/JHEP04(2017)078}
[\arxiv{1702.05489} [hep-th]].

\bibitem{Baron:2018lve}
W.~H.~Baron, E.~Lescano and D.~Marqu\'es,
``The generalized Bergshoeff-de Roo identification,''
JHEP \textbf{11} (2018), 160
\doi{10.1007/JHEP11(2018)160}
[\arxiv{1810.01427} [hep-th]].

\bibitem{Marques:2015vua}
D.~Marques and C.~A.~Nunez,
``T-duality and \ensuremath{\alpha}'-corrections,''
JHEP \textbf{10} (2015), 084
\doi{10.1007/JHEP10(2015)084}
[\arxiv{1507.00652} [hep-th]].

\bibitem{Bedoya:2014pma}
O.~A.~Bedoya, D.~Marques and C.~Nunez,
``Heterotic $\alpha$'-corrections in Double Field Theory,''
JHEP \textbf{12} (2014), 074
\doi{10.1007/JHEP12(2014)074}
[\arxiv{1407.0365} [hep-th]].

\bibitem{Ortin:2020xdm}
T.~Ort\'{\i}n,
``O(n, n) invariance and Wald entropy formula in the Heterotic Superstring
 effective action at first order in $\alpha'$,''
JHEP \textbf{01} (2021), 187
\doi{10.1007/JHEP01(2021)187}
[\arxiv{2005.14618} [hep-th]].




  


































\end{thebibliography}
\end{document}